\documentclass[aps,twocolumn,showpacs,superscriptaddress,pre]{revtex4}
\usepackage{amsmath}
\usepackage{amssymb}
\usepackage{mathtext}
\usepackage{graphicx}
\usepackage{color}

\begin{document}

\newcommand {\wX} {\widetilde{X}}
\newcommand {\ph} {\varphi}
\newcommand {\Th} {\varTheta}
\newcommand {\la} {\langle}
\newcommand {\ra} {\rangle}
\newcommand {\lla} {\left\langle}
\newcommand {\rra} {\right\rangle}
\newcommand {\calK} {\mathcal{K}}

\title[Multicomponent gases bubbly horizon in porous media \& Temperature wave]{Fertilization of liquid-saturated porous medium with multicomponent gases\\ due to surface temperature oscillation}

\author{Anastasiya V.\ Dolmatova}
\affiliation{Institute of Continuous Media Mechanics, UB RAS,
             Academician Korolev Street 1, 614013 Perm, Russia}
\affiliation{Institute for Information Transmission Problems, RAS,
             Bolshoy Karetny Pereulok 19, 127051 Moscow, Russia}
\author{Denis S.\ Goldobin}
\affiliation{Institute of Continuous Media Mechanics, UB RAS,
             Academician Korolev Street 1, 614013 Perm, Russia}

\date{\today}

\begin{abstract}
We study non-isothermal diffusive transport of a weakly-soluble two-component substance in a liquid-saturated porous medium being in contact with the reservoir of this substance. Particular attention is given to the example case of infiltration of nitrogen and oxygen from the atmosphere under the annual temperature oscillation.
The surface temperature of the porous medium half-space oscillates in time, which results in a decaying solubility wave propagating deep into the porous medium.
In such a system, the zones of saturated solution and nondissolved phase coexist with the zones of undersaturated solution; these zones migrate with time. Moreover, the solubility of a multicomponent substance depends on its composition, which results in a much more intricate mathematical model of solubility as compared to the single-component case.
% The effect is firstly considered for the case of annual oscillation of the surface temperature of water-saturated ground being in contact with atmosphere.
We describe the phenomenon of formation of a near-surface bubbly horizon due to the temperature oscillation. An analytical theory of the phenomenon is developed.
For multicomponent solutions we report the formation of diffusion boundary layer, which is not possible for single-component solutions. We construct an analytical theory for this boundary layer; in particular, we derive effective boundary conditions for the problem of the diffusive transport beyond this layer.
%Further, the treatment is extended to the case of higher frequency oscillations and case of weakly-soluble solids and liquids.

 % end of PACS codes

\end{abstract}

\maketitle

\section{Introduction}
In many geological and technological systems, non-convective mechanisms of gas transport through a porous medium play a decisive role~\cite{Li-Orr-Benson-2021,diff-b-l,Davie-Buffett-2001,Haacke-Westbrook-Riley-2008}.
The diffusive transport in bubbly media~\cite{Haacke-Westbrook-Riley-2008,Donaldson-etal-1997-1998,Goldobin-Brilliantov-2011,Krauzin-Goldobin-2014,Goldobin-Krauzin-2015,Maryshev-Goldobin-2018} and media with condensed nondissolved phase~\cite{Haacke-Westbrook-Riley-2008,Davie-Buffett-2001,Goldobin-CRM-2013,Goldobin-etal-EPJE-2014} exhibits nontrivial features; these features become even more intricate under nonisothermal conditions. In these systems the presence of the nondissolved phase keeps the local solute concentration being equal to the solubility. Hence, the solute concentration is not a `free' variable, but it becomes a function of the local pressure and temperature. In its turn, a nonzero divergence of the diffusion flux of the solute, driven by the concentration gradient, does not change the local solute concentration; instead, it redistributes the mass of the nondissolved (gaseous or condensed) phase~\cite{Goldobin-Brilliantov-2011}. In this way, the dynamics of the systems with a nondissolved phase are governed by new effects and mechanisms, which are not in line with the intuition gained with the diffusion dynamics in undersaturated solutions. The role of these effects and mechanisms becomes especially pronounced for the systems, where the nondissolved phase is immobilized (for instance, it is trapped in a porous medium) and solubility is small~\cite{Goldobin-Brilliantov-2011}. For the immobilized nondissolved phase, the only transport mechanism is the diffusion transfer via the solute; and when the solubility is small, the mass stored in a nondissolved phase can be by orders of magnitude larger than the mass contained in the solution.

In Ref.~\cite{Krauzin-Goldobin-2014}, the impact of the surface temperature oscillation, which creates a solubility wave, on the diffusion transfer was studied for a porous medium with everywhere-present nondissolved phase. The systems, where the zones of nondissolved phase coexist with the zones of undersaturated solution, exhibit more rich and sophisticated dynamics~\cite{Goldobin-Krauzin-2015,Maryshev-Goldobin-2018}. A fluid-saturated porous medium being in contact with a reservoir of a weakly-soluble substance ({\it e.g.}, atmosphere) is an example of such a system. In theoretical work~\cite{Goldobin-Krauzin-2015}, the solubility wave was revealed to lead to the formation of a near-surface bubbly horizon. In this bubbly horizon, the mass of the guest substance exceeds the time-average solubility at the surface, meaning fertilization of the porous massif. An analytical theory explaining the mechanism of this phenomenon was constructed and yielded a decent quantitative agreement with numerical simulation.

The contact with a reservoir/atmosphere is mathematically represented via the conditions at the boundary, where we assume the instantaneous solute concentration to be equal to the solubility and no nondissolved phase. Owing to this boundary condition, for a single-component guest substance, the substance amount in the near-surface part of the horizon of the nondissolved phase is as high as the maximal-over-period solubility~\cite{Goldobin-Krauzin-2015}. This excess over the time-averaged solubility results in an enhanced fertilization of the porous medium with the guest substance.
However, for a two-component guest substance the situation changes. In this paper we show that a narrow diffusion boundary layer emerges, where the amount of the guest substance is shifted towards the time-averaged solubility. Thus, the effect of the fertilization enhancement by the temperature oscillation is somewhat reduced. This diffusion boundary layer is impossible for a single-component guest substance, and the mechanism of its formation differs from the ones for ordinary diffusion boundary layers of gas in geological porous media~\cite{diff-b-l}. This boundary layer forms as a result of the difference in the diffusive mobilities of solute components. Beyond the boundary layer, the time-averaged behavior of the bulk of the nondissolved phase horizon is qualitatively similar to that for the single-component case.

The phenomenon under consideration can influence the systems with diverse origin of the temperature oscillation, including technological systems: filters, nuclear and chemical reactors, underground $\mathrm{CO_2}$-burial systems, {\it etc.} However, for the sake of definiteness, we focus our study on the case of a two-component gas in the presence of the hydrostatic pressure gradient, which is important for geological systems. Mathematically, in the case where nondissolved phase is solid or liquid, the solubility becomes nearly independent of pressure. Thus, the theory we construct can be technically extended to this case by setting the pressure gradient to zero. In the text below, we keep our consideration as general as possible without the necessity to extend the paper content drammatically: we explicitly give the reference to ``gases'' or ``bubbles'' where the statement is correct only for the case of gases with a strong hydrostatic pressure gradient and ``nondissolved phase'' where the statement can be extended to solids and liquids in a straightforward way.

An enhanced fertilization of sediments by the atmospheric gases creates more favorable conditions for local flora and fauna and influences geochemical processes. For natural deposits of methane hydrate in seabed sediments, the impact of temperature waves on the deposit and the gas release from it is of interest in connection to the Glacial--Interglacial cycles~\cite{iceage} and potential scenarios of climate change~\cite{Hunter-etal-2013}.

The paper is organized as follows. In Sec.~\ref{sec2} we construct the mathematical model for the mass transfer in two-component gas solutions, where the zones of nondissolved phase intermingle with the zones of undersaturated solution and the solubility depends on the composition of the nondissolved phase. In Sec.~\ref{sec22} the system dynamics is illustrated with numerical simulations. The framework for the analytical theory, based on the separation of the time scales for heat and mass diffusion, is introduced in Sec.~\ref{sec3}. In Sec.~\ref{sec4}, the theory of diffusion boundary layer is constructed and the effective boundary conditions for the system immediately beyond this layer are derived. In Sec.~\ref{sec5}, the theory of the formation of the horizon of the nondissolved phase beyond the boundary layer is constructed. The results are summarized in Conclusion, Sec.~\ref{sec6}.

\section{Diffusion in saturated multi-component solutions}
\label{sec2}
\subsection{Physical and mathematical model}
\label{sec21}
It is convenient to consider the problem we address in terms of the molar solute concentration which is the molar amount of solute per 1 mole of solvent.
For a single-component perfect gas, in thermodynamic equilibrium, the molar solute concentration $X_s^{(0)}$ in contact with the gaseous phase---solubility---is determined by the Henry's law~\cite{Henry-1803}:
\begin{equation}
P=K_HX_s^{(0)}\,,
\label{eq-pmm01}
\end{equation}
where $P$ is pressure and $K_H$ is the Henry's law constant. According to the scaled particle theory~\cite{Pierotti-1976}, one can write
\begin{equation}
 K_H=\frac{P_0}{X_s^{(0)}(T_0,P_0)}
\frac{T}{T_0}
 \exp\left[q\left(\frac{1}{T_0}-\frac{1}{T}\right)\right]\,,
\label{eq-pmm02}
\end{equation}
where $T_0$ and $P_0$ are reference values, the choice of which is guided merely by convenience,  $X^{(0)}(T_0,P_0)$ is the solubility at the reference temperature and pressure; the parameter $q\equiv-G_i/k_\mathrm{B}$, with $G_i$ being the interaction energy between a solute molecule and the surrounding solvent molecules and $k_\mathrm{B}$ being the Boltzmann constant, is provided in Table~\ref{params} for several typical gases. The scaled particle theory allows calculating $q$ and $X^{(0)}(T_0,P_0)$ from first principles, while Eq.~(\ref{eq-pmm02}) is more general; with empirically determined $q$ and $X^{(0)}(T_0,P_0)$, Eq.~(\ref{eq-pmm02}) is valid for moderate temperature variation and pressure values for which the gas can be treated as a perfect one (that is typically up to several tens of atmospheres).

For multi-component gases, each gas component in the solution creates the partial pressure $P_j$ in the gaseous phase according to
\begin{equation}
P_j=K_{H,j}(T)\,X_{s,j}\,,
\label{eq-pmm03}
\end{equation}
where $K_{H,j}$ is the Henry's law constant of the specie $j$ and $X_{s,j}$ is the concentration of the solution of specie $j$. With the molar fraction $Y_j$ of specie $j$ in the gaseous phase, partial pressure  \[
P_j=P\,Y_j
\]
and pressure $P=\sum_jP_j$, {\it i.e.}, $\sum_jY_j=1$.

Under standard conditions, the solubility of typical gases (see Table~\ref{params}) is so much small, that if the dissolved molecules form gas bubbles, the volumetric fraction of these bubbles in pore fluid will be negligibly small. Hence, it is convenient to quantify the composition of the pore fluid with $X_{s,j}$, $X_{b,j}$, and $X_{\Sigma,j}=X_{s,j}+X_{b,j}$, where $X_{s,j}$ is the number of the molecules of specie $j$ in the solution divided by the total number of molecules in the liquid and gaseous phases, $X_{b,j}$ is the number of the molecules of specie $j$ in the gaseous phase (bubbles) divided by the total number of molecules, and $X_{\Sigma,j}$ is the net molar fraction of specie $j$ in the pore fluid. Since the volumetric fraction of the gaseous phase in pores is small, $X_{s,j}$ nearly equals the molar solute concentration, and, in what follows, we neglect the quantitative discrepancy between $X_{s,j}$ and the solute concentration.

\subsubsection{Solubility of two-component gas}
Henceforth we consider a two-component gas, which is also a reasonable model for the Earth's atmosphere, where nitrogen and oxygen comprise $99\%$ of molar composition. With a given content of the pore fluid, $X_{\Sigma,1}$ and $X_{\Sigma,2}$, one can evaluate whether the gaseous phase forms, and calculate the composition of the solution and gaseous phase when the latter appears. For the gaseous phase to be formed, the maximal solute concentrations $\max(X_{s,j})=X_{\Sigma,j}$ should be sufficient to create the net vapour pressure $\max(P_1+P_2)$ exceeding pressure $P$; according to Eq.~(\ref{eq-pmm03}), the {\em condition of formation of the gaseous phase} reads
\begin{equation}
K_{H,1}X_{\Sigma,1}+K_{H,2}X_{\Sigma,2}>P\,.
\label{eq-pmm04}
\end{equation}
When the gaseous phase forms, its equilibrium composition is determined by Eq.~(\ref{eq-pmm03}),
\[
K_{H,j}X_{s,j}=P\,Y_j\,,
\]
and relations
\[
X_{s,j}+X_{b,j}=X_{\Sigma,j}\,,
\]
\[
X_{b,1}/X_{b,2}=Y_1/Y_2\,,
\]
\[
Y_1+Y_2=1\,.
\]
These 6 equations (with $j=1,2$) compose the equation system for 6 unknown variables $X_{s,j}$, $X_{b,j}$, $Y_j$ with $j=1,2$. This equation system possesses unique physically meaningful solution:
\begin{eqnarray}
&&
X_{s,1}=2X_{\Sigma,1}X_{s,1}^{(0)}\bigg(
 X_{\Sigma,1}+X_{\Sigma,2}+X_{s,1}^{(0)}-X_{s,2}^{(0)}
\nonumber\\[7pt]
&&\qquad\qquad
 {}+\Big[\Big(X_{\Sigma,1}-X_{\Sigma,2}-X_{s,1}^{(0)}+X_{s,2}^{(0)}\Big)^2
\nonumber\\[7pt]
&&\qquad\qquad\qquad\qquad
 {}+4X_{\Sigma,1}X_{\Sigma,2}\Big]^{1/2}
 \bigg)^{-1},
\label{eq-pmm05}
\\[10pt]
&&
X_{s,2}=X_{s,2}^{(0)}\left(1-X_{s,1}/X_{s,1}^{(0)}\right)\,,
\label{eq-pmm06}
\end{eqnarray}
where $X_{s,j}^{(0)}\equiv P/K_{H,j}$ is the solubility of a single component gas. Solution (\ref{eq-pmm05})--(\ref{eq-pmm06}) is physically meaningful when condition (\ref{eq-pmm04}) is fulfilled.
With condition (\ref{eq-pmm04}) and Eqs.~(\ref{eq-pmm05}) and (\ref{eq-pmm06}), one can calculate the local equilibrium state of the system for given $X_{\Sigma,1}$ and $X_{\Sigma,2}$.

\subsubsection{Temperature and pressure fields}
Geological systems are typically much more uniform in the horizontal directions than in the vertical one. Hence, we restrict our consideration to the one-dimensional case; the system is assumed to be homogeneous in the horizontal directions. We assume the $z$-axis to be oriented downwards and its origin to be on the porous medium surface.

We focus our study on the effect of the surface temperature oscillation on the system. We consider harmonic oscillation, $T_0+\Theta_0\cos{\omega t}$, where $T_0$ is the mean temperature, $\Theta_0$ is the oscillation amplitude, $\omega$ is the temperature oscillation cyclic frequency. In particular, annual oscillations of surface temperature only slightly deviate from their harmonic reduction (e.g., see~\cite{Yershov-1998}). The heat diffusion equation
 $\partial T/\partial t=\chi\mathrm{\Delta}T$
with no-heat-flux condition deep below the surface (at infinity) and imposed surface temperature yields
\begin{equation}
T(z,t)=T_0+\Theta_0e^{-kz}\cos(\omega t-kz)\,,
\quad
k=\sqrt{\omega/2\chi}\,,
\label{eq-pmm07}
\end{equation}
where $\chi$ is the heat diffusivity and $z$ is the distance from the surface of porous medium. The pressure field is a hydrostatic one:
\begin{equation}
P=P_0+\rho gz\,,
\label{eq-pmm08}
\end{equation}
where $P_0$ is the atmospheric pressure, $\rho$ is the liquid density, and $g$ is the gravity acceleration.

%%%%%%%%%%%%%%%%%%%%%%%%%%%%%%%%%%%%%%%%%%%%%%%%%%%%%%%%%%%%%%%%%%%%%%%
\begin{table}[t]
\caption{Chemical physical properties of solutions of nitrogen, oxygen, methane and carbon dioxide in water. Equations (\ref{eq-pmm01})--(\ref{eq-pmm02}) with $q$ and $X^{(0)}(T_0,P_0)$ specified in the table fit the experimental data from~\cite{solubility}. Equation (\ref{eq-pmm12}) with provided values of effective radius $R_d$ and parameter $\nu$ of the solute molecules fits the experimental data from~\cite{diffusion}.}
\begin{center}
\begin{tabular}{cp{0.7cm}p{0.7cm}p{0.7cm}p{0.55cm}}
\hline\hline
 & $\mathrm{N_2}$ & $\mathrm{O_2}$ & $\mathrm{CH_4}$ & $\mathrm{CO_2}$
 \\
\hline
$q=-G_i/k_\mathrm{B}$ (K)
 & 781 & 831 & 1138 & 1850 \\[5pt]
$X^{(0)}(20^\circ\mathrm{C},1\,\mathrm{atm})$ ($10^{-5}$)
 & 1.20 & 2.41 & 2.60 & 68.7 \\[5pt]
$R_d$ ($10^{-10}\,\mathrm{m}$)
 & 1.48 & 1.29 & 1.91 & 1.57 \\[5pt]
$\nu$ ($10^{-5}\,\mathrm{Pa\cdot s}$)
 & 9.79 & 16.3 & 28.3 & 4.68 \\[3pt]
\hline\hline
\end{tabular}
\end{center}
\label{params}
\end{table}
%%%%%%%%%%%%%%%%%%%%%%%%%%%%%%%%%%%%%%%%%%%%%%%%%%%%%%%%%%%%%%%%%%%%%%%

\subsubsection{Diffusion transport equations}
Since the nondissolved phase is immobilised in pores, the mass is transferred solely by molecular diffusion through the intersticial liquid and governed by equations
\begin{equation}
\frac{\partial X_{\Sigma,j}}{\partial t}
 =\frac{\partial}{\partial z}\left(D_j\frac{\partial}{\partial z} X_{s,j}\right),
\label{eq-pmm09}
\end{equation}
where $D_j$ is the effective molecular diffusion coefficient of specie $j$. Compared to the molecular diffusion coefficients in bulk of pure liquid, say $D_{\mathrm{mol},j}$, the effective coefficients are influenced by the pore network geometry (tortuosity) and the adsorption of the diffusing agents on porous matrix. On the time scales of our interest the adsorption does not lead to anomalous diffusion; it only changes the effective rate of normal diffusion~\cite{Gregg-Sing-1982}. Although the importance of thermal diffusion effect~\cite{Bird-Stewart-Lightfoot-2007} was demonstrated for gases~\cite{Goldobin-Brilliantov-2011} and methane
hydrate~\cite{Goldobin-CRM-2013,Goldobin-etal-EPJE-2014} on geological time scales, for the system of our interest it can be neglected~\cite{Krauzin-Goldobin-2014}. The solute concentrations $X_{s,j}$ are determined by Eqs.~(\ref{eq-pmm05})--(\ref{eq-pmm06}) where condition (\ref{eq-pmm04}) is fulfilled ({\it i.e.}, the gaseous phase forms), and equal to the net molar fraction $X_{\Sigma,j}$, otherwise. In the latter case, $X_{b,j}=0$.

In this mathematical model the dissolution process (as well as the opposite process of formation of the nondissolved phase from the solution) occurs much faster than the change in the temperature field and the diffusive redistribution of the solute mass. In real systems, the dissolution time scales for solid nondissolved phase are assessed as hours~\cite{Buffett-Zatsepina-2000}, which is small compared to the time scales of temperature oscillation and diffusive transport. For gases the dissolution process is even faster. The hysteresis effects possible for some phase transformations in narrow pore channels~\cite{Anderson-Tohidi-Webber-2000} are also neglected in our study.

Eq.\,(\ref{eq-pmm09}) is accurate for the case where macroscopic porosity is spatially uniform and the nondissolved phase occupies a negligible fraction of the pore volume, which holds true for the systems under consideration.

At the upper boundary we assume contact with the atmosphere, which means that $Y_j=Y_{j0}$, where $Y_{j0}$ is the molar fraction of specie $j$ in atmosphere, and, hence,
\begin{equation}
X_{s,j}(z=0,t)=\frac{Y_{j0}P_0}{K_{H,j}\big(T(z=0,t)\big)}\,.
\label{eq-pmm10}
\end{equation}
Deep below the surface we assume the no-flux condition and the absence of the nondissolved phase;
\begin{equation}
\left.\frac{\partial X_{s,j}}{\partial z}\right|_{z=+\infty}=0\,,
\qquad
X_{b,j}(z=+\infty)=0\,.
\label{eq-pmm11}
\end{equation}
Note that two boundary conditions are required at $z\to+\infty$; however, due to the specificity of our system, one boundary condition, Eq.\,(\ref{eq-pmm10}), is sufficient at $z=0$. Indeed, since $X_{s,j}(z=0)$ are never less than solubility, the value of $X_{b,j}$ at the point $z=0$ does not influence the system dynamics; the condition for it is redundant.

Generally, all material properties of the system depend on temperature and pressure. However, feasible relative variations of the absolute temperature are small. Hence, one can neglect variation of those parameters which depend on temperature polynomially and consider variation of only those parameters which depend on temperature exponentially: the latter parameters are the Henry's law constants (\ref{eq-pmm02}) and the molecular diffusion coefficients $D_j$. The only parameter sensitive to pressure is the gas solubility [see Eq.~(\ref{eq-pmm03})].

We employ the following dependence of molecular diffusion on temperature~\cite{Bird-Stewart-Lightfoot-2007};
\begin{equation}
D_{\mathrm{mol},j}(T)=\frac{k_\mathrm{B}T}{2\pi\mu R_{d,j}}
 \frac{\mu+\nu_j}{2\mu+3\nu_j}\,,
\label{eq-pmm12}
\end{equation}
where $\mu$ is the dynamic viscosity of the solvent, $R_{d,j}$ is the effective radius of the  molecules of solute $j$ with the ``coefficient of sliding friction'' $\beta_j$, $\nu_j=R_{d,j}\beta_j/3$. The dependence of dynamic viscosity on temperature can be described by a modified Frenkel formula~\cite{Frenkel-1955}
\begin{equation}
\mu(T)=\mu_0\exp\frac{a}{T+\tau}\,.
\label{eq-pmm13}
\end{equation}
For water, the coefficient
 $\mu_0=2.42\cdot10^{-5}\,\mathrm{Pa\cdot s}$,
 $a=W/k_\mathrm{B}=570\,\mathrm{K}$
($W$ is the activation energy) and $\tau=-140\,\mathrm{K}$. For the effective diffusion coefficient $D_j$ the relative variation with temperature is assumed to be the same as for $D_{\mathrm{mol},j}$. The parameter values for aqueous solutions of typical gases are provided in Table~\ref{params}.

%%%%%%%%%%%%%%%%%%%%%%%%%%%%%%%%%%%%%%%%%%%%%%%%%%%%%%%%%
\begin{figure*}[!t]
\centerline{
{\sf (a)}\hspace{-12pt}
\includegraphics[width=0.22\textwidth]%
 {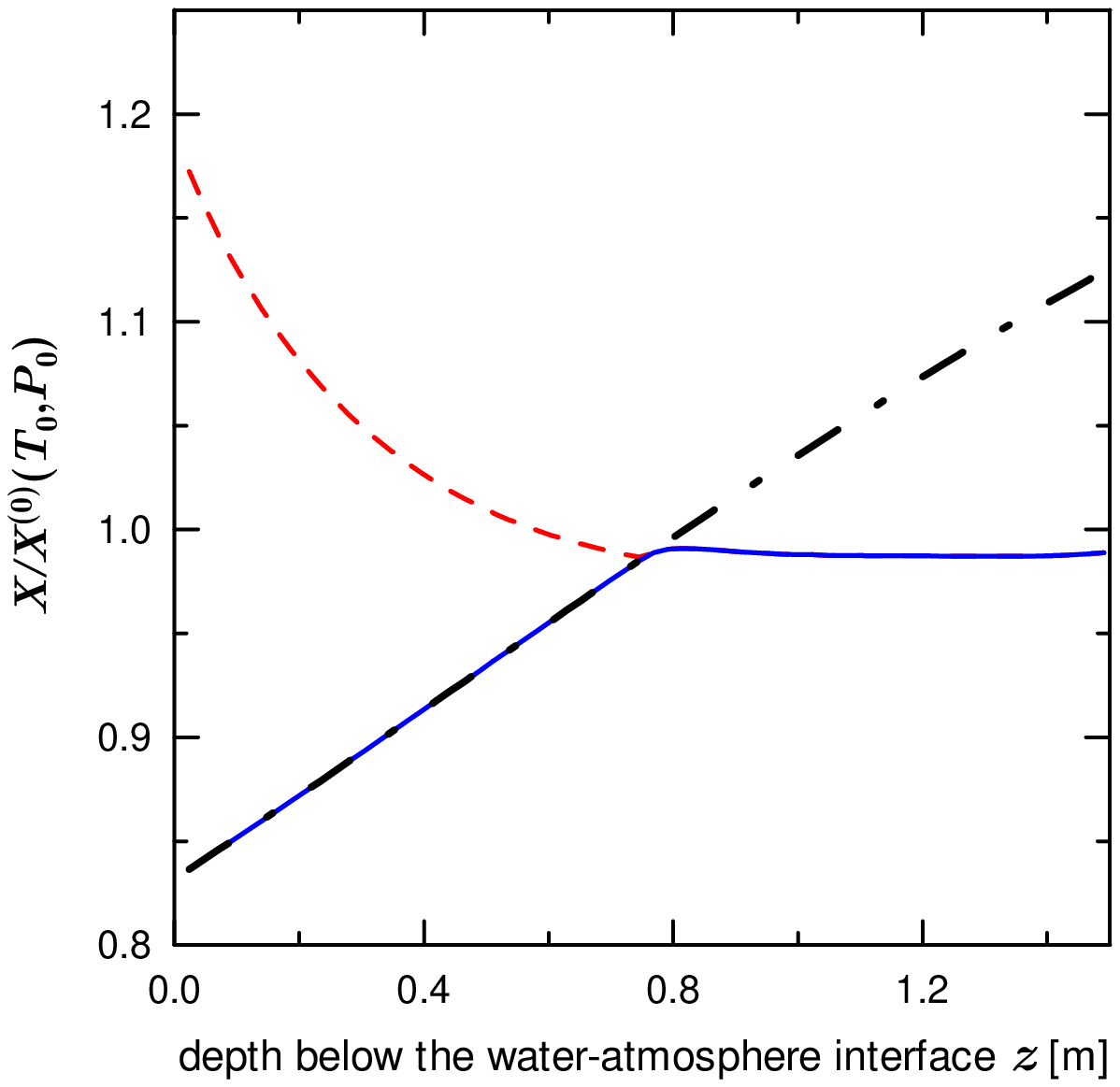}
\quad
{\sf (b)}\hspace{-12pt}
\includegraphics[width=0.22\textwidth]%
 {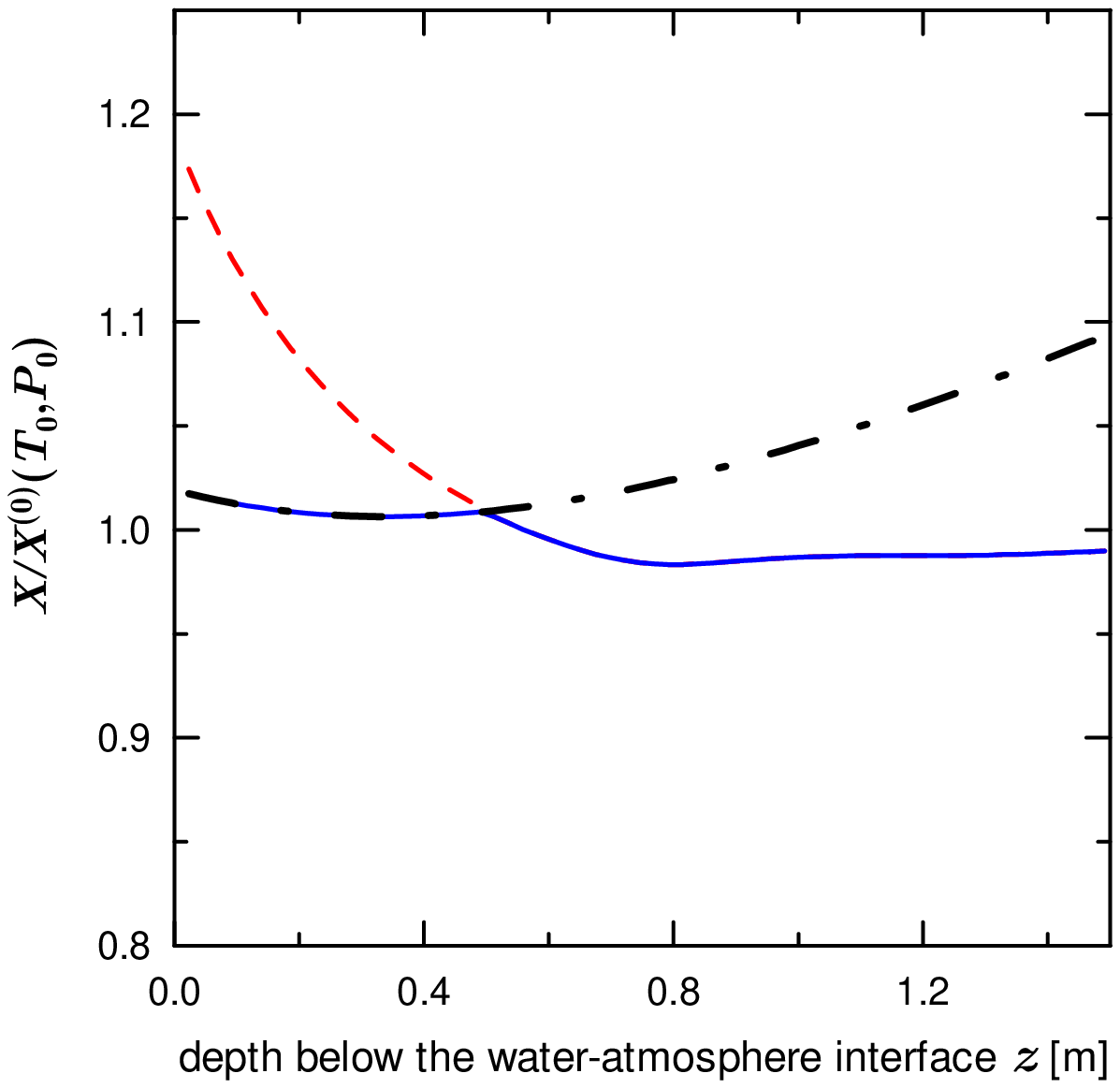}
\quad
{\sf (c)}\hspace{-12pt}
\includegraphics[width=0.22\textwidth]%
 {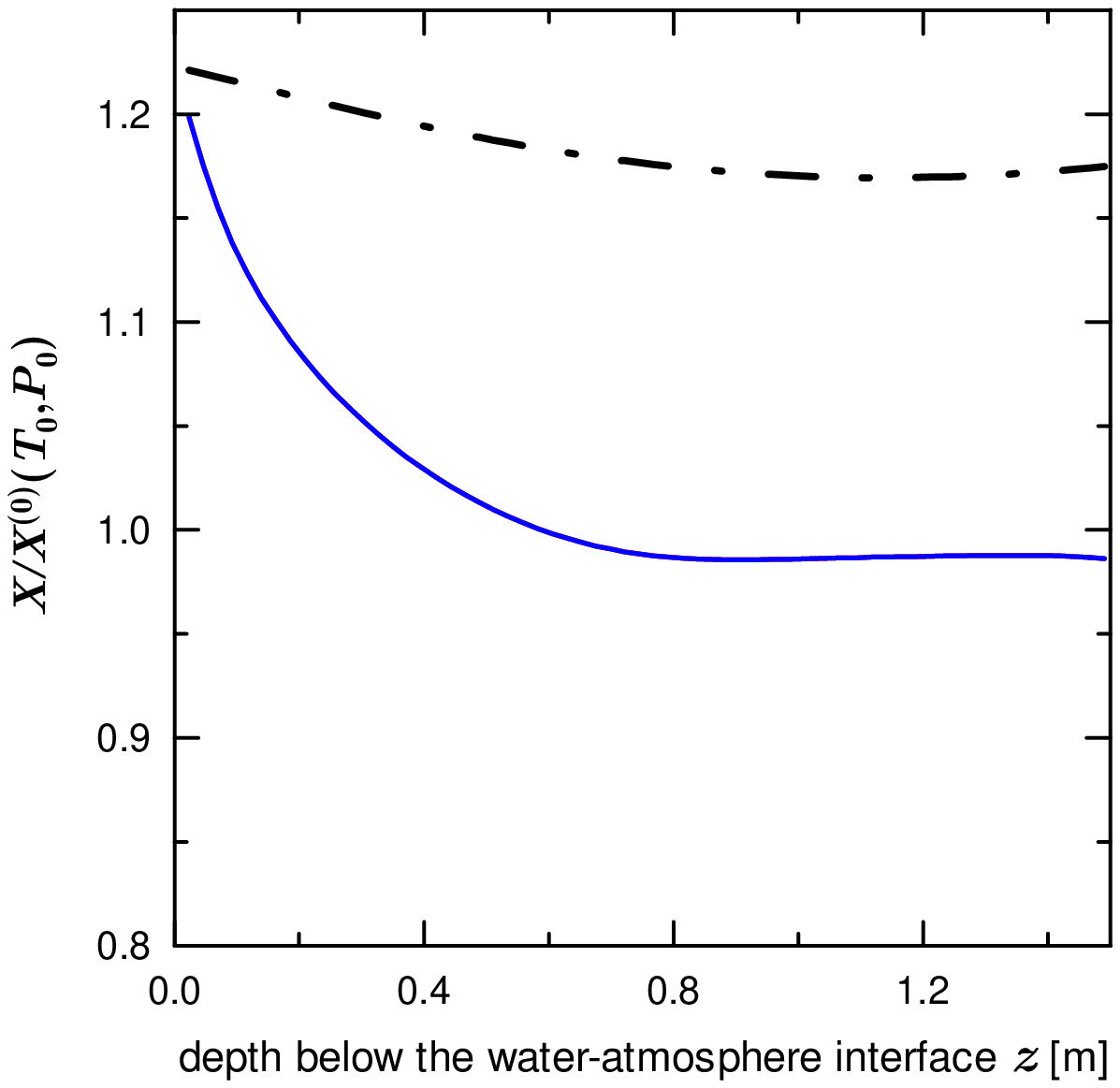}
\quad
{\sf (d)}\hspace{-12pt}
\includegraphics[width=0.22\textwidth]%
 {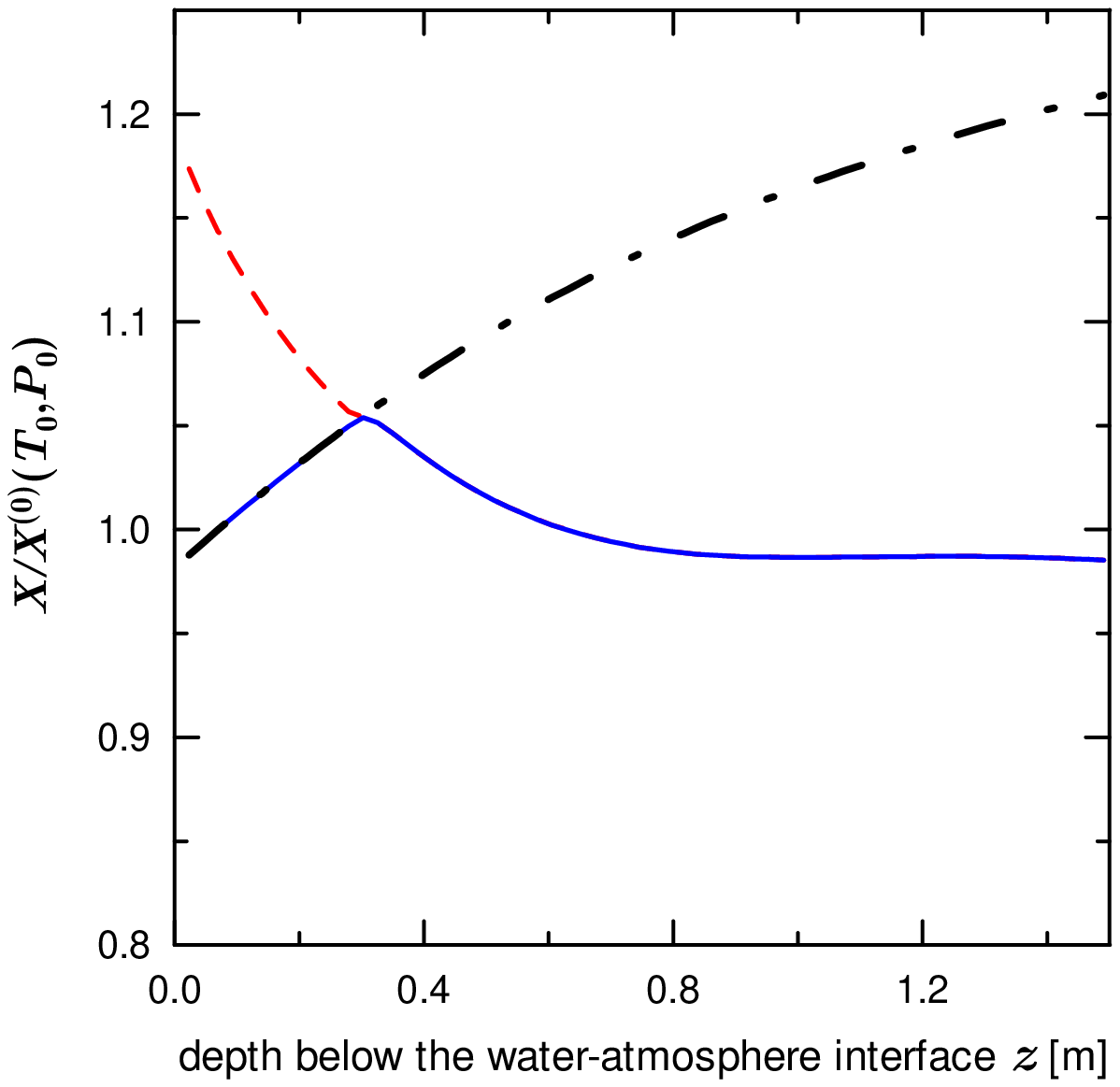}
}

  \caption{(Color online) Snapshots of the oscillating solubility profile $X^{(0)}$ of nitrogen for the annual temperature wave in water-saturated ground and the single-component atmosphere are plotted with the black dash-dotted lines for different phases of surface temperature oscillation: (a)~midsummer, $\varphi(z=0)=0$, (b)~midautumn, $\varphi(z=0)=\pi/2$, (c)~midwinter, $\varphi(z=0)=\pi$, (d)~midspring,  $\varphi(z=0)=3\pi/2$. The blue solid lines represent the solution molar concentration $X_s$. The red dashed lines show the net molar fraction $X_\Sigma$ of nitrogen molecules in the pore fluid where the bubbly phase is present. The bubbly fraction $X_b$ is given by the difference between the red dashed and blue solid profiles (it does not exist for the cold winter period, when solubility is high). Parameters: $T_0=300\,\mathrm{K}$, $\Theta_0=15\,\mathrm{K}$.
 }
  \label{fig1}
\end{figure*}
%%%%%%%%%%%%%%%%%%%%%%%%%%%%%%%%%%%%%%%%%%%%%%%%%%%%%%%%%

%%%%%%%%%%%%%%%%%%%%%%%%%%%%%%%%%%%%%%%%%%%%%%%%%%%%%%%%%
\begin{figure*}[!t]
\centerline{
{\sf (a)}\hspace{-12pt}
\includegraphics[width=0.22\textwidth]%
 {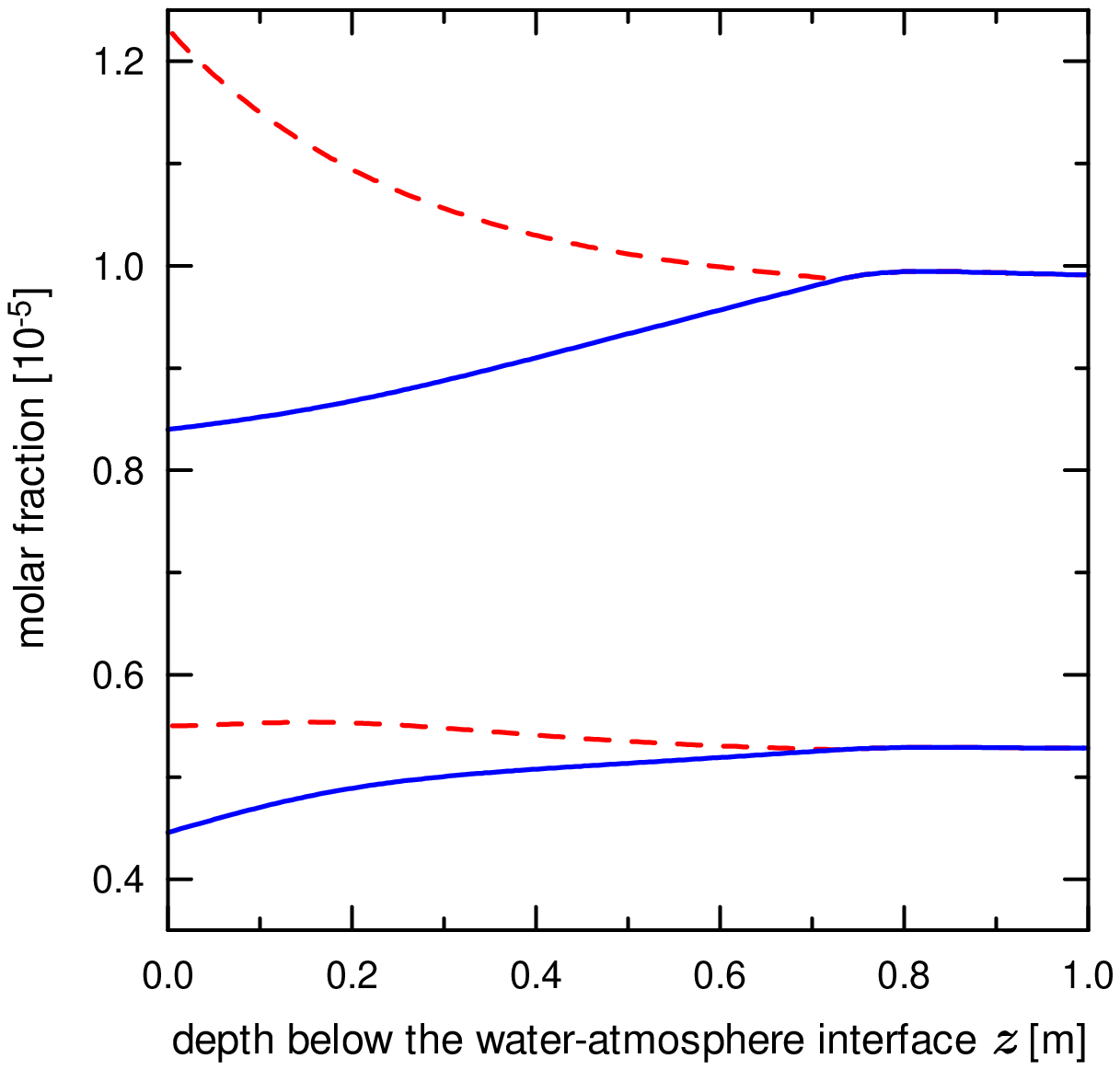}
\quad
{\sf (b)}\hspace{-12pt}
\includegraphics[width=0.22\textwidth]%
 {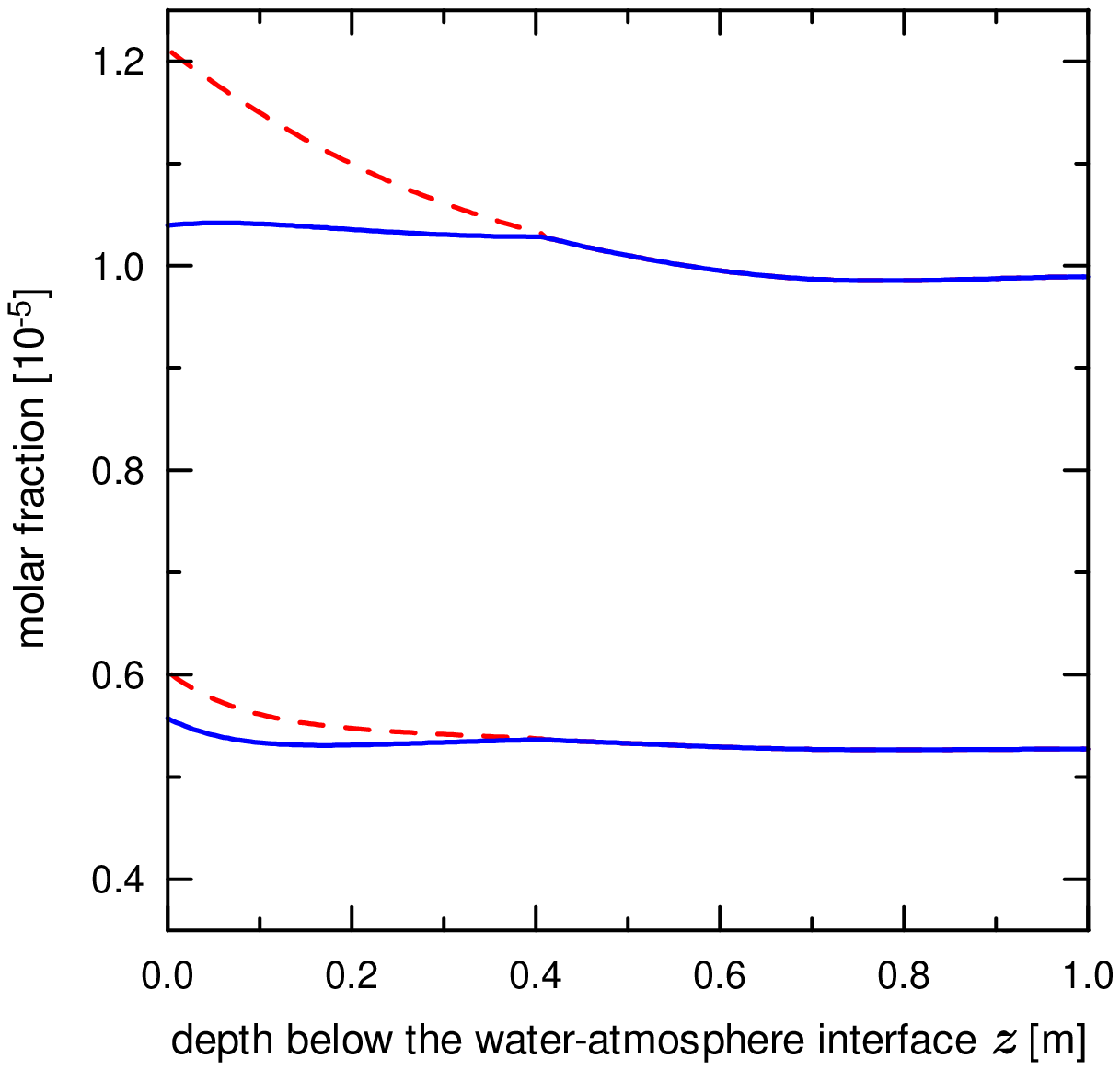}
\quad
{\sf (c)}\hspace{-12pt}
\includegraphics[width=0.22\textwidth]%
 {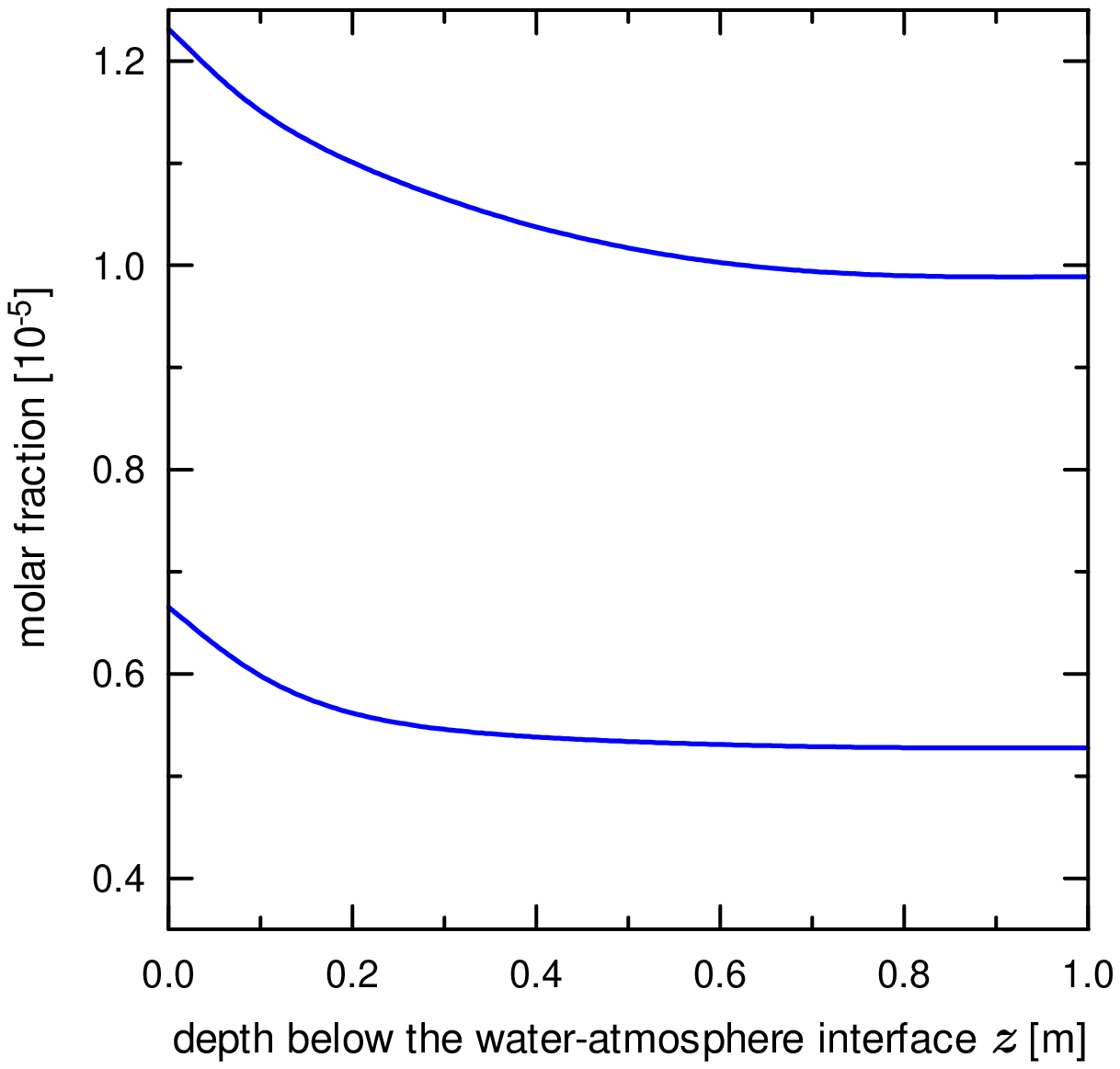}
\quad
{\sf (d)}\hspace{-12pt}
\includegraphics[width=0.22\textwidth]%
 {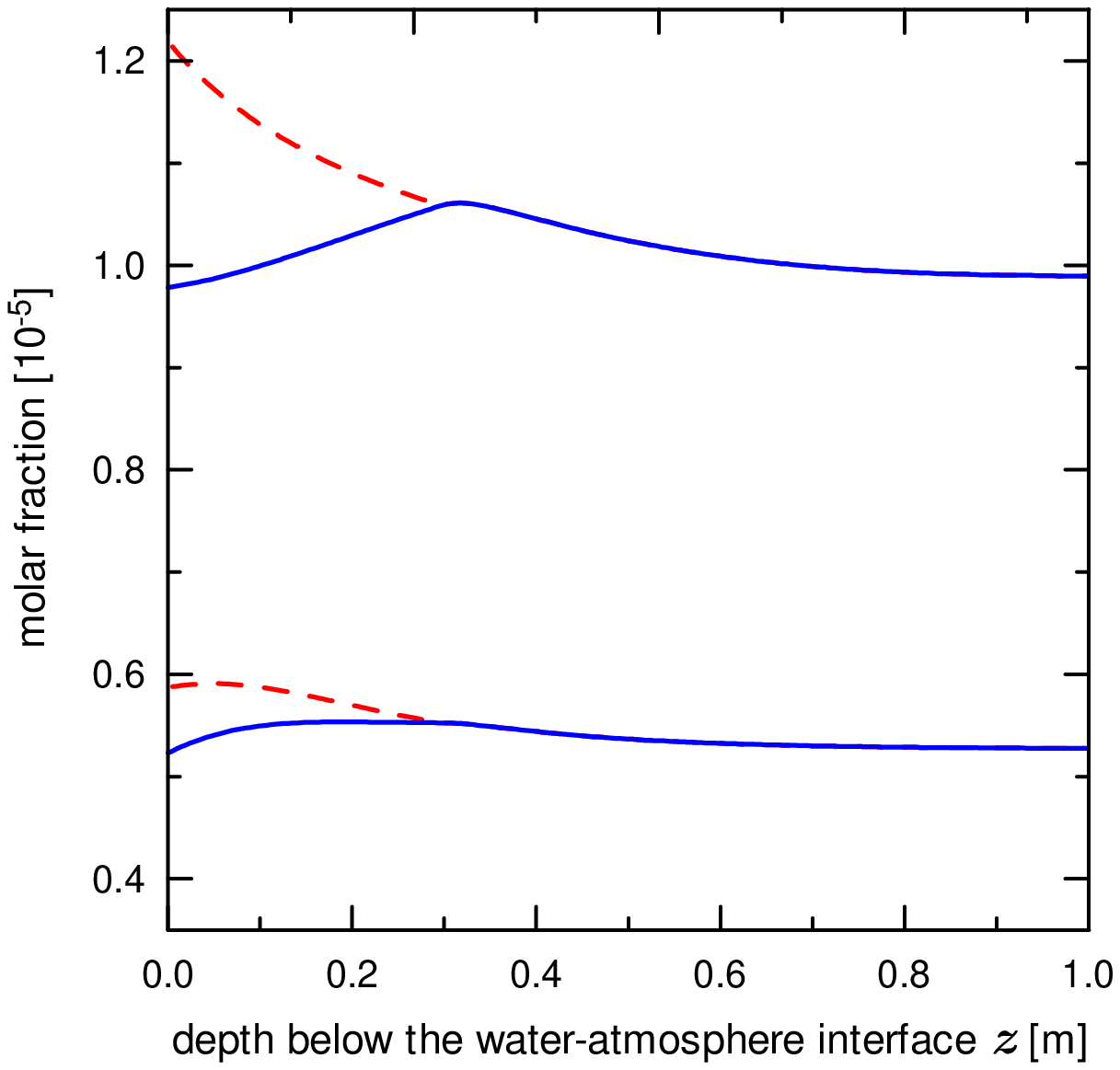}
}

  \caption{(Color online) Snapshots of the oscillating profiles of the solute molar concentration of nitrogen $X_{s,1}$ (upper) and oxygen $X_{s,2}$ (lower) for the annual temperature wave in water-saturated ground and the two-component atmosphere are plotted with the blue solid lines for different phases of surface temperature oscillation: (a)~$\varphi(z=0)=0$, (b)~$\varphi(z=0)=\pi/2$, (c)~$\varphi(z=0)=\pi$, (d)~$\varphi(z=0)=3\pi/2$. The red dashed lines show the net molar fractions $X_{\Sigma,1}$ and $X_{\Sigma,2}$ of nitrogen and oxygen molecules in the pore fluid where the bubbly phase is present. The bubbly fraction $X_b$ is given by the difference between the red dashed and blue solid profiles. Parameters: $T_0=300\,\mathrm{K}$, $\Theta_0=15\,\mathrm{K}$.
 }
  \label{fig2}
\end{figure*}
%%%%%%%%%%%%%%%%%%%%%%%%%%%%%%%%%%%%%%%%%%%%%%%%%%%%%%%%%

%%%%%%%%%%%%%%%%%%%%%%%%%%%%%%%%%%%%%%%%%%%%%%%%%%%%%%%%%
\begin{figure*}[!t]
\centerline{
{\sf (a)}\hspace{-12pt}
\includegraphics[width=0.22\textwidth]%
 {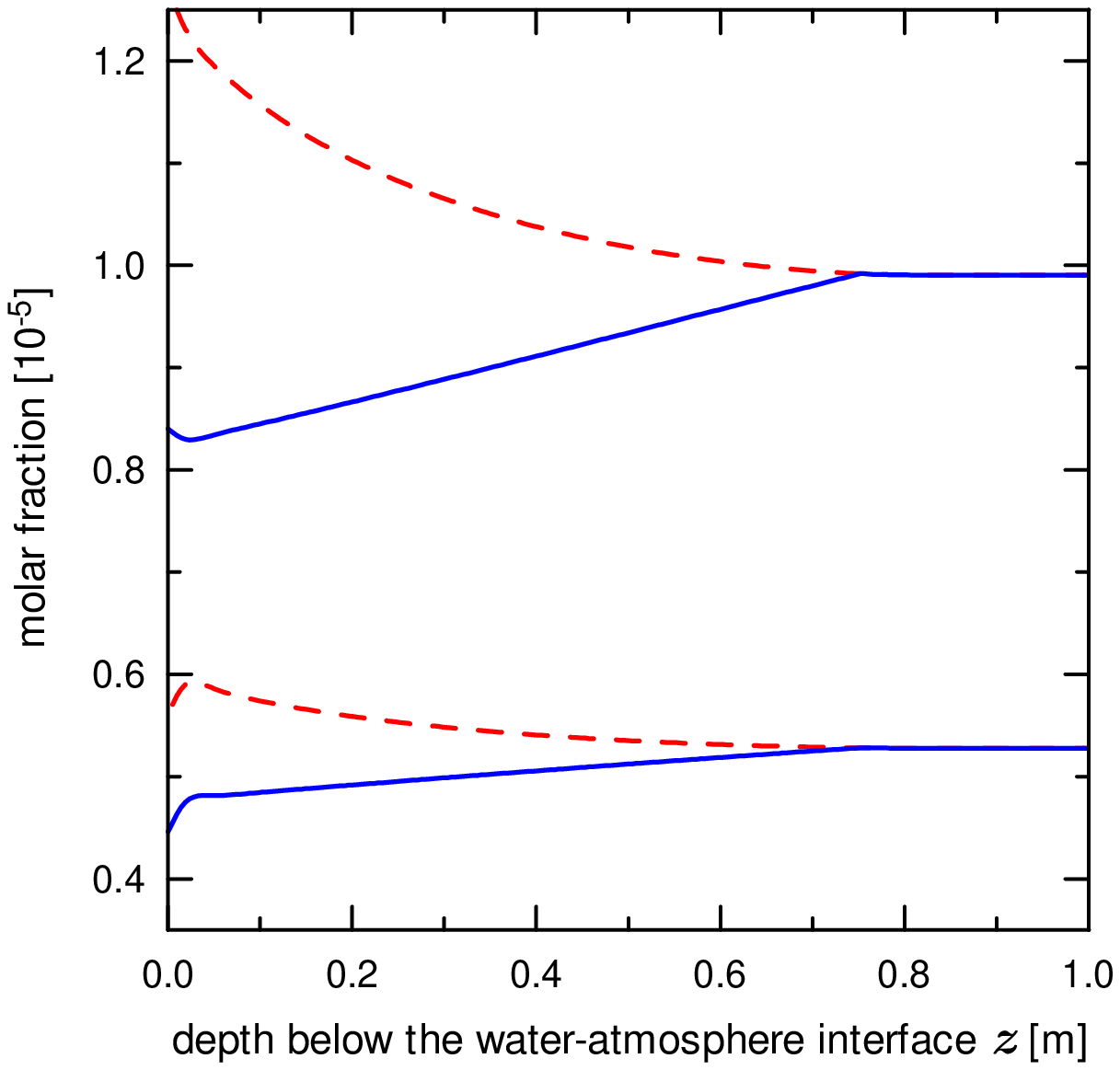}
\quad
{\sf (b)}\hspace{-12pt}
\includegraphics[width=0.22\textwidth]%
 {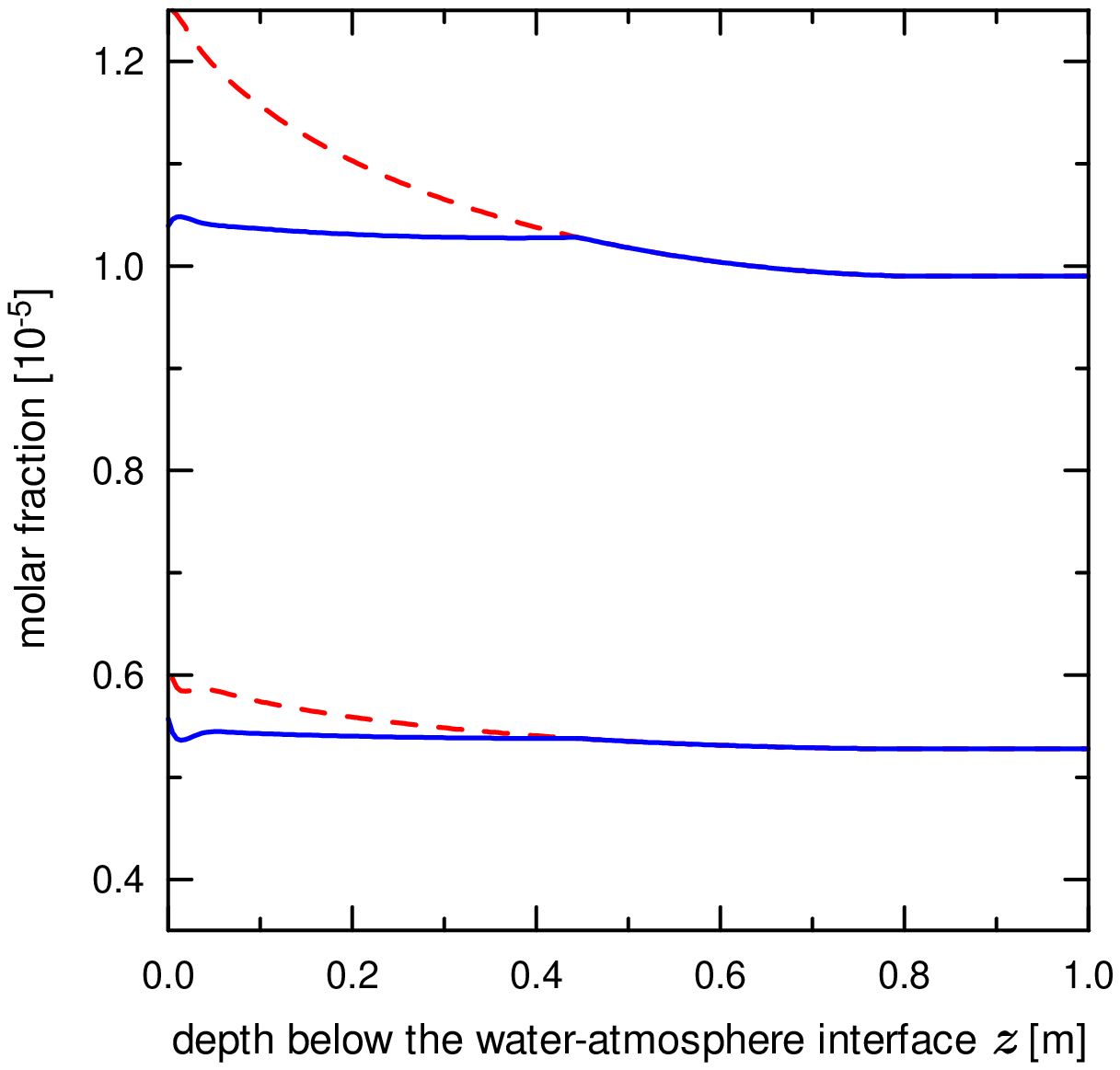}
\quad
{\sf (c)}\hspace{-12pt}
\includegraphics[width=0.22\textwidth]%
 {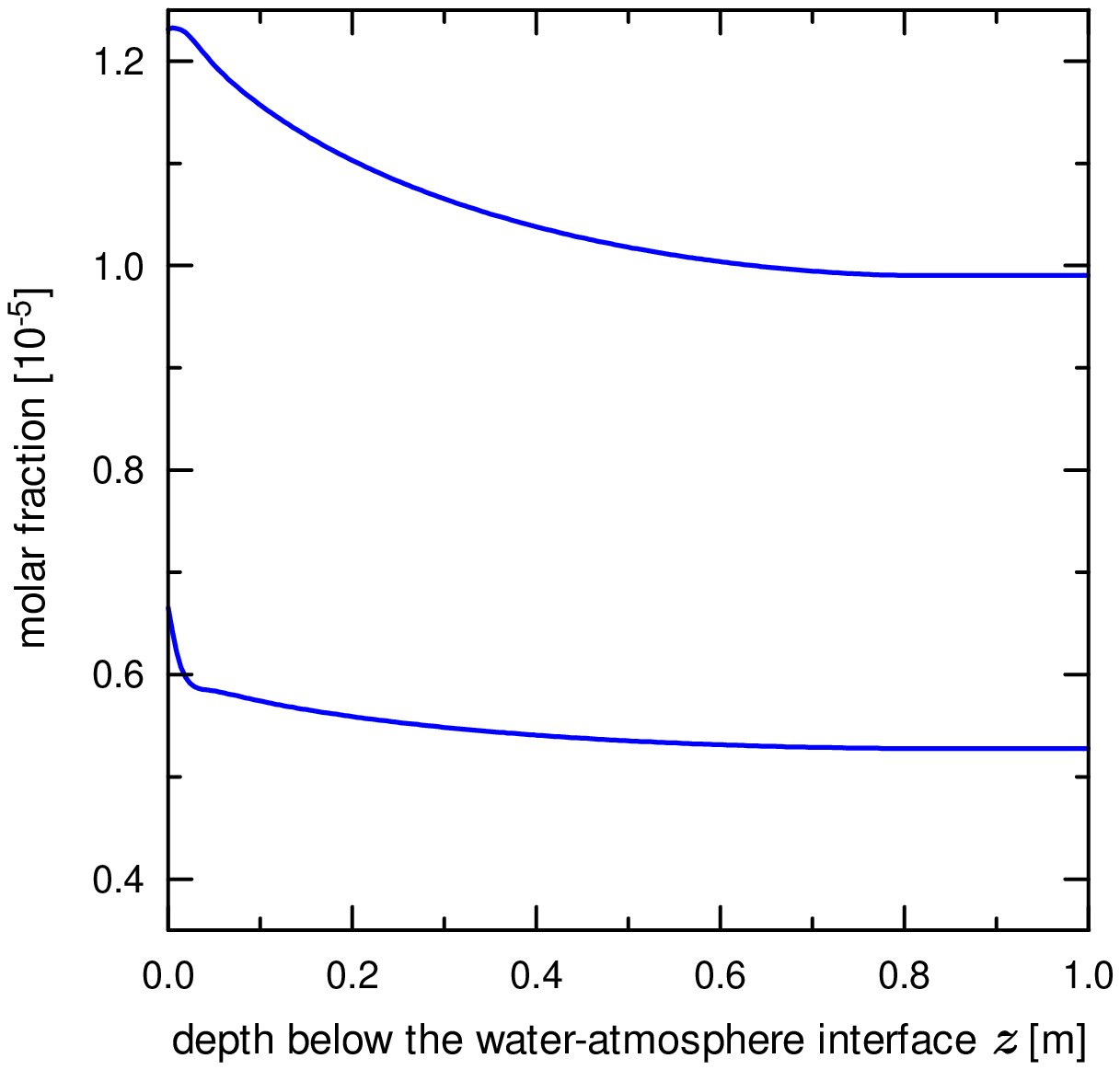}
\quad
{\sf (d)}\hspace{-12pt}
\includegraphics[width=0.22\textwidth]%
 {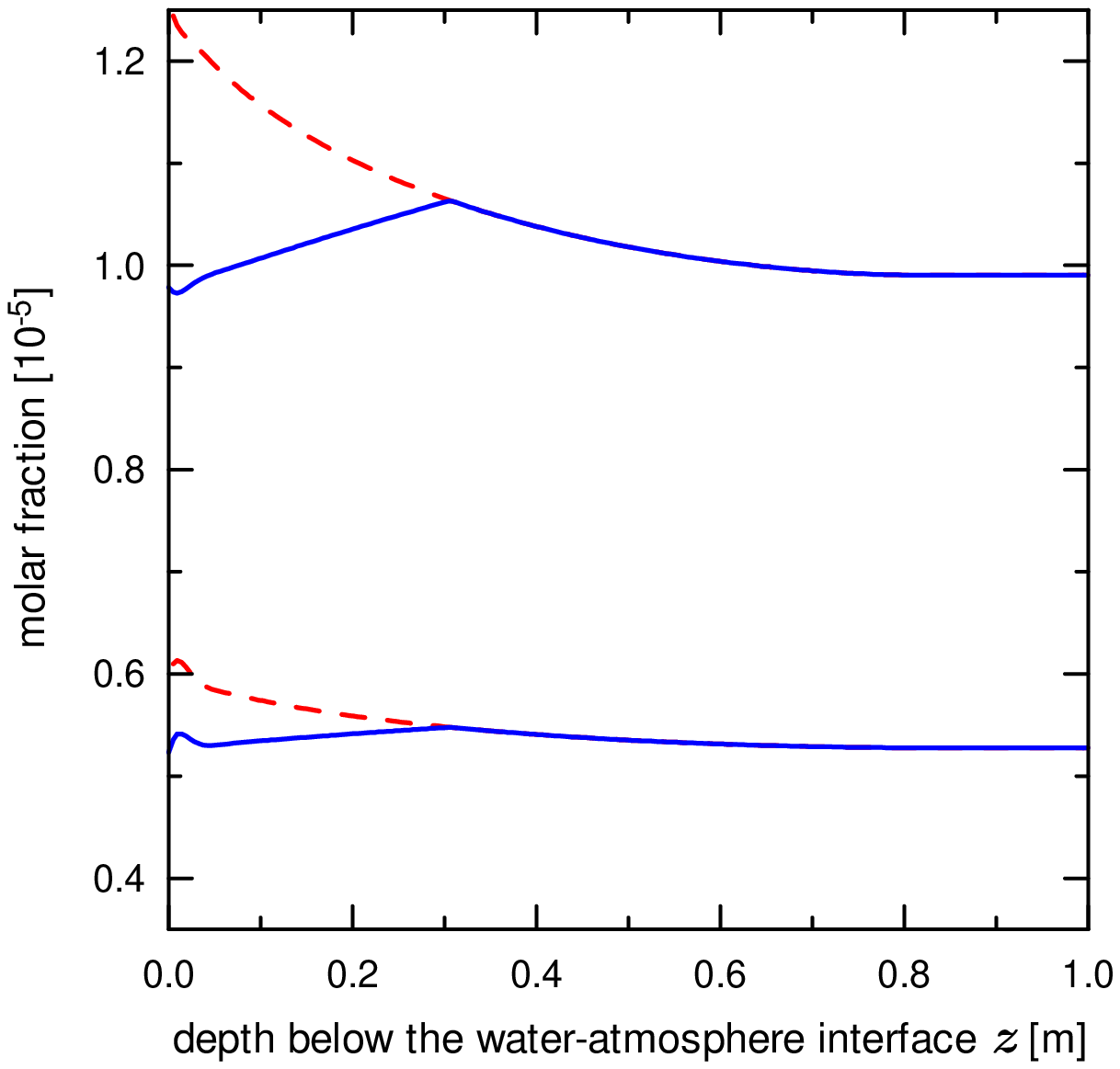}
}

  \caption{(Color online) Oscillating profiles of the molar concentration of nitrogen and oxygen for the annual temperature wave in water-saturated ground and the two-component atmosphere are plotted for the case of a clogged porous matrix with low effective diffusion coefficient $D_j=0.01D_{\mathrm{mol},j}$.
  See Caption to Fig.~\ref{fig2} for description; parameters: $T_0=300\,\mathrm{K}$, $\Theta_0=15\,\mathrm{K}$.
 }
  \label{fig3}
\end{figure*}
%%%%%%%%%%%%%%%%%%%%%%%%%%%%%%%%%%%%%%%%%%%%%%%%%%%%%%%%%

%%%%%%%%%%%%%%%%%%%%%%%%%%%%%%%%%%%%%%%%%%%%%%%%%%%%%%%%%
\begin{figure}[!t]
\centerline{
\includegraphics[width=0.375\textwidth]%
 {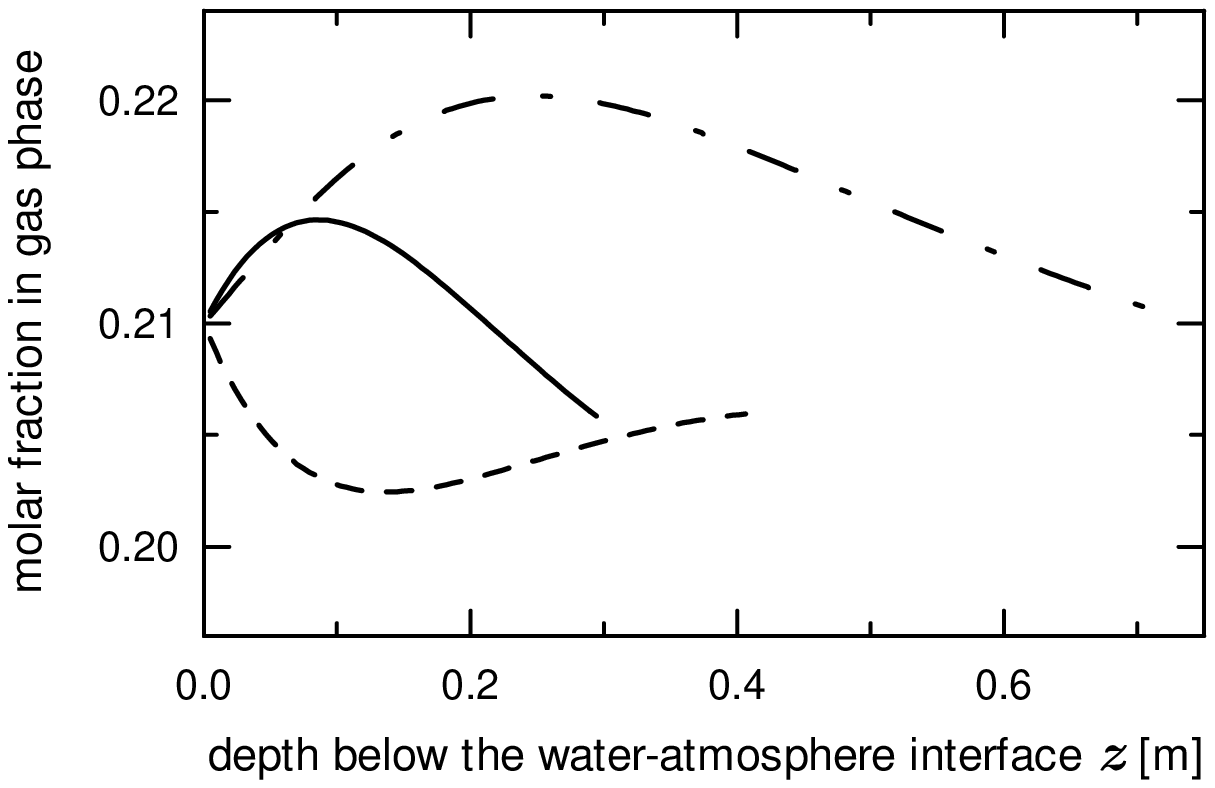}
}

  \caption{The variation of the composition of the gaseous phase caused by the different diffusive mobility of components is plotted for the case presented in Fig.~\ref{fig2}. Dash-dotted line: midsummer, $\varphi(z=0)=0$; dashed line: midautumn, $\varphi(z=0)=\pi/2$; solid line: midspring, $\varphi(z=0)=3\pi/2$; no gaseous phase for $\varphi(z=0)=\pi$.
 }
  \label{fig4}
\end{figure}
%%%%%%%%%%%%%%%%%%%%%%%%%%%%%%%%%%%%%%%%%%%%%%%%%%%%%%%%%

\subsection{Numerical simulation}
\label{sec22}
Numerical simulation was performed for the nitrogen--oxygen atmosphere as follows.
The $z$-coordinate was discretized so that the zone of the penetration of the nondissolved phase, say $L$, was represented by $200$ nodes. Tests proved that the oscillations of the solution components weakly penetrate beyond this zone and exponentially decay there. Hence, we adopted the condition of no diffusion flux and no nondissolved phase at the depth $2L$ as an accurate approximation of the condition at infinity, Eq.~(\ref{eq-pmm11}). At a given time step we calculated the fields of the Hery's law constants~(\ref{eq-pmm02}) for two gas components from the instantaneous temperature field~(\ref{eq-pmm07}). At the surface ($z=0$) the fields $X_{s,1}$ and $X_{s,2}$ were set equal to the current solubility of the respective components according to Eq.~(\ref{eq-pmm03}) with $P_j=PY_j$ and $Y_j$ given by the atmosphere composition. In all the other mesh nodes, for the current fields $X_{\Sigma,1}$ and $X_{\Sigma,2}$ the condition~(\ref{eq-pmm04}) of the presence of the nondissolved phase was checked (in each node). Where the nondissolved phase is present, we calculated the solute concentrations $X_{s,1}$ and $X_{s,2}$ with Eqs.~(\ref{eq-pmm05}) and (\ref{eq-pmm06}); otherwise, we set $X_{s,1}=X_{\Sigma,1}$ and $X_{s,2}=X_{\Sigma,2}$. Further, employing the central difference scheme of Eq.~(\ref{eq-pmm09}), we calculated the fields $X_{\Sigma,1}$ and $X_{\Sigma,2}$ for the next time step from the current fields $X_{s,1}$ and $X_{s,2}$.

For the illustration of the mechanisms of system dynamics let us first consider the results of numerical simulation for a simplified case of a single-component atmosphere composed solely by nitrogen and subject to the annual temperature oscillation. The modelling shows that for all initial conditions, after a transient process, the system arrives to a unique stable time-periodic regime presented in Fig.~\ref{fig1}.

The linear growth of the solubility with depth, created by the hydrostatic pressure gradient, is modulated by the decaying temperature wave~(\ref{eq-pmm07}). The oscillating solubility profile~(\ref{eq-pmm01})--(\ref{eq-pmm02}) for the temperature wave~(\ref{eq-pmm07}) and pressure~(\ref{eq-pmm08}) is plotted in Fig.~\ref{fig1} with the black dash-dotted curve. The oscillations of the solubility profile create a nearly frozen profile of the molar fraction $X_\Sigma(z)$. The molar fraction of the matter in the nondissolved phase $X_b(z)$ is the difference between profiles $X_\Sigma(z)$ and $X_s(z)$. Profile $X_\Sigma(z)$ almost attains the maximal solubility (for the minimal temperature---midwinter) close to the surface, $z=0$; here the nondissolved phase exists almost always, except for a short coldest time interval of the cycle. Profile $X_\Sigma(z)$ monotonously decreases with depth, along with the span of the timeinterval when the nondissolved phase is present at $z$, down to the depth where the latter is never formed. Below this depth $X_\Sigma(z)=X_s(z)$ is nearly uniform and only slightly perturbed during the year cycle. The heterogeneity of profile $X_\Sigma(z)$ rapidly decays with depth in this zone. The asymptotic value $X_\infty$ turns out to be very close to the annual-mean gas solubility at the surface.

The profile of the net molar fraction is nearly constant during the oscillation period as, for typical liquids, the molecular diffusion coefficient is by three orders of magnitude smaller than the heat diffusion coefficient; therefore, the diffusive redistribution of mass is a slow process against the background of fast waves of temperature and solubility, which is a function of temperature. Such a strong separation of time scales lends the opportunity to develop an analytical theory of the system dynamics. This theory can elucidate the generic mechanisms of the formation of the nondissolved phase horizon and aids in the interpretation of the results of numerical simulation. In Secs.~\ref{sec3}--\ref{sec5}, we construct the theory for both one- and two-component guest substances.

In the case of two-component guest substance the behavior becomes more complicated (Figs.~\ref{fig2}, \ref{fig3}), since the solubilities of components differently depend on temperature and thus respond to its wave, and the components themselves diffuse with different rate. The most striking manifestation of this complication is the formation of a diffusion boundary layer in a thin near-surface zone of the porous medium, which can be clearly seen in Fig.~\ref{fig3}. In fact, this layer also impacts the system dynamics in Fig.~\ref{fig2}. However, the spatial scale separation between the diffusion boundary layer $\delta_\mathrm{diff}=\sqrt{2D/\omega}$ and the temperature boundary layer (the zone of penetration of the temperature wave) $\delta_T=\sqrt{2\chi/\omega}$ differ only by factor $\delta_T/\delta_\mathrm{diff}=\sqrt{\chi/D}\sim 30$ in Fig.~\ref{fig2}, and the bubbly horizon occupies only nonlarge part of the temperature boundary layer. Hence, the diffusion boundary layer is less than by an order of magnitude smaller that the bubbly zone; a naked eye cannot distinguish this layer against the background of other nonstationarities of the solubility profiles in Fig.~\ref{fig2}. However, below we will present the evidence that the theory of Secs.~\ref{sec3}--\ref{sec5} assuming the presence of a diffusion boundary layer here is in perfect agreement with observed profiles. In Fig.~\ref{fig3}, where the diffusive transport is impaired by the pore clogging---typical for consolidated sediments---and $D_j=0.01D_{\mathrm{mol},j}$, the scale $\delta_\mathrm{diff}\propto\sqrt{D_j}$ becomes by one order of magnitude smaller; here one can clearly recognize this layer.

The second visible phenomenon, oscillations of the solute composition, is more obvious and intuitively expected. In Fig.~\ref{fig4}, the variation of the solute composition from Fig.~\ref{fig2} is presented with high resolution.
With disabled diffusive transport, in the presence of the nondissolved phase, the change of temperature would result in the equilibrium redistribution of component molecules between the solute and nondissolved phases.
The magnitude of the composition variation is primarily contributed by the difference in the relative variation of solubility parameters, $K_{H,j}(T)$, with the variation of temperature. On top of that, the slow diffusive transport disperses these variations in space. While the temperature dependence of $K_{H,j}(T)$ creates variations of the solute concentrations, the local mass of components is not changed by these variations themselves; however, the diffusive dispersion of these variations already causes the mass redistribution. Generally, this mass transfer also creates a nonzero average flux. To summarize, the instantaneous variations in the solubility profiles, like the ones in Fig.~\ref{fig4}, are controlled by the dissimilarity of functions $K_{H,1}(T)/K_{H,1}(T_0)$ and $K_{H,2}(T)/K_{H,2}(T_0)$ and practically not affected by a slow diffusion, but the long-term net mass transfer is proportional to the diffusion coefficient. In Fig.~\ref{fig4}, the variation hardly attains 1 percent point as the reference difference
\[
\frac{\left.\frac{\mathrm{d}K_{H,1}(T)}{\mathrm{d}T}\right|_{T_0}\Theta_0}{K_{H,1}(T_0)} -\frac{\left.\frac{\mathrm{d}K_{H,2}(T)}{\mathrm{d}T}\right|_{T_0}\Theta_0}{K_{H,2}(T_0)}
\]
is small for nitrogen and oxygen even for $\Theta_0=15\,\mathrm{K}$ and $T_0=300\,\mathrm{K}$.

\section{Analytical theory}
\label{sec3}
For constructing analytical theory we assume small temperature oscillations. It is more productive to start from the basic physical equations and utilize the smallness of certain quantities in the course of derivation than to deal with Eqs.~(\ref{eq-pmm05})--(\ref{eq-pmm06}) and simplify them for the case of small oscillations.

We consider two-component gas bubbles in liquid under the hydrostatic pressure gradient and non-isothermal conditions. Local-equilibrium partial pressure $P_j$ in the gaseous phase being in contact with the solution of the specie $j$, the molar concentration of which is $X_{s,j}$, is
\begin{equation}
P_j=K_{H,j}(T)\,X_{s,j},
\label{eq-m01}
\end{equation}
where $K_{H,j}$ is the Henry's law constant of the specie $j$. With oscillating temperature $T=T_0+\Theta_0\cos{\omega t}$ of the sediment--atmosphere interface, the temperature field in sediments is
\begin{eqnarray}
T(z,t)=T_0+\Theta(z,t)=T_0+\Theta_0 e^{-kz}\cos(\omega t-kz)\nonumber\\[5pt]
=T_0+\Theta_0 e^{-kz}\cos\varphi\,,\qquad
\label{eq-m02}
\end{eqnarray}
where temperature oscillation phase $\varphi=\omega t-kz$, $k=\sqrt{\omega/2\chi}$\,, $z$ is the depth below the sediment--atmosphere interface, $\chi$ is the heat diffusivity coefficient. Hence, the local-equilibrium solute concentrations $X_{s,j}$ in the presence of bubbles of a two-component gas obey
\begin{eqnarray}
K_{H,1}X_{s,1}+K_{H,2}X_{s,2}=P_0(1+bz)\,,
\label{eq-m03}
\end{eqnarray}
where $P_0$ is the atmospheric pressure, $b=\rho_lg/P_0$, $\rho_l$ is the liquid density, $g$ is the gravity. We will use notation
\[
K_j=K_{H,j}/P_0\,,
\]
and linearize the dependence of Henry's law constants on temperature;
\begin{equation}
K_j=K_{j0}\big(1+a_j\Theta+\mathcal{O}\big((a_j\Theta)^2\big)\big)\,,
\label{eq-m04}
\end{equation}
where
\[
K_{j0}\equiv K_{j}(T_0)
\]
and
\[
a_j\equiv\frac{1}{K_j}\left(\frac{\partial K_j}{\partial T}\right)_{T_0}\,.
\]
The ratio of molecule numbers in gaseous phase $X_{b,1}/X_{b,2}=P_1/P_2$, and, according to Eq.~(\ref{eq-m01}),
\[
\frac{X_{\Sigma,1}-X_{s,1}}{X_{\Sigma,2}-X_{s,2}}=\frac{K_1X_{s,1}}{K_2X_{s,2}}\,.
\]
When the relative variations of solubility are non-large, on the left-hand side of the latter equality, one find the ratio of small numbers, while on the r.h.s., there is the ratio of non-small numbers, values of which are slightly perturbed by the variation of $K_j$ and pressure. Hence, one can approximately assume constancy of this ratio, which means the constancy of the composition of the gaseous phase, as observed in numerical simulation (the variations in Fig.~\ref{fig4} are below 1\%);
\begin{equation}
\frac{X_{\Sigma,1}-X_{s,1}}{X_{\Sigma,2}-X_{s,2}}\approx\frac{Y_{10}}{Y_{20}}\,,
\label{eq-m05}
\end{equation}
where
\[
Y_{j0}=K_{j0}X_{s,j0}
\]
is the molar fraction of specie $j$ in atmosphere; notice,
\[
Y_{10}+Y_{20}=1\,.
\]

Let us consider the deviation of the gas mass distribution in the interstitial fluid from the no-temperature-oscillation state
\[
X_{\Sigma,j}|_{\Theta_0=0}=X_{s,j}|_{\Theta_0=0}=X_{s,j0}=\frac{Y_{j0}}{K_{j0}}\,.
\]
Specifically,
\[
X_{\Sigma,j}=X_{s,j0}+\widetilde{X}_{\Sigma,j}\,,
\]
\[
X_{s,j}=X_{s,j0}+\widetilde{X}_{s,j}\,.
\]
In terms of $\widetilde{X}_{\Sigma,j}$ and $\widetilde{X}_{s,j}$, Eq.~(\ref{eq-m03}) reads (to the leading order)
\begin{eqnarray}
K_{10}\widetilde{X}_{s,1}+K_{20}\widetilde{X}_{s,2}=bz-\Theta\sum_j a_jY_{j0}\,,
%+\mathcal{O}(K_{j0}a_j\Theta\widetilde{X}_{s,j})
\label{eq-m06}
\end{eqnarray}
Eq.~(\ref{eq-m05}) reads
\begin{eqnarray}
Y_{20}\widetilde{X}_{s,1}-Y_{10}\widetilde{X}_{s,2}
 =Y_{20}\widetilde{X}_{\Sigma,1}-Y_{10}\widetilde{X}_{\Sigma,2}\,.
\label{eq-m07}
\end{eqnarray}
Eqs.~(\ref{eq-m06})--(\ref{eq-m07}) form a self-contained equation system for $\wX_{s,j}$ as functions of $\Theta$ and $\wX_{\Sigma,j}$. Note, Eqs.~(\ref{eq-m06})--(\ref{eq-m07}) are valid for the sediment domain, where the net amount $X_{\Sigma,j}$ is sufficient for formation of the gaseous phase, {\it i.e.}, according to Eq.~(\ref{eq-m06}), $\Theta>\Theta(\varphi_\ast)$ determined by the condition
\begin{eqnarray}
K_{10}\widetilde{X}_{\Sigma,1}+K_{20}\widetilde{X}_{\Sigma,2}=bz-\Theta(\varphi_\ast)\sum_j a_jY_{j0}\,.
\label{eq-m08}
\end{eqnarray}
Where $\Theta<\Theta(\varphi_\ast)$, all the guest gas molecules are dissolved and $\wX_{s,j}=\wX_{\Sigma,j}$.

Transport of guest molecules operates through the liquid phase via molecular diffusion of the solution. Due to the smallness of the ratio of the molecular and thermal diffusion coefficients for liquids, the net concentration profiles $X_{\Sigma,j}$ are nearly frozen on the time scale of one period of temperature oscillation. Hence, it is enough to calculate the period-average molecular diffusion flux. In the porous medium domain where bubbles appear for some part of the oscillation period,
\begin{eqnarray}
\langle{J_j}\rangle =\frac{1}{t_p}\int\limits_{t_1}^{t_2}
 \left(-D_j\frac{\partial\wX_{s,j}}{\partial z}\right)dt
\qquad\qquad\nonumber\\[5pt]
{}+\frac{1}{t_p}\int\limits_{t_2}^{t_1+t_p}\left(-D_j\frac{\partial\wX_{\Sigma,j}}{\partial z}\right)dt\,.
\label{eq-m09}
\end{eqnarray}
Here $t_p=2\pi/\omega$ is the oscillation period, $t_1<t_2$ are the time instants between which local temperature is high enough, $\Theta>\Theta(\varphi_\ast)$, so that not all amount of guest gas molecules can be dissolved and the solution flux is driven by the solute concentration gradient $d\wX_{s,j}/dz$, determined by equation system (\ref{eq-m06})--(\ref{eq-m07}). For the rest of the period, $\wX_{s,j}=\wX_{\Sigma,j}$ and the solute flux is driven by the gradient of the net concentration of the guest molecules. Eq.~(\ref{eq-m09}) can be rewritten in terms of the temperature oscillation phase $\varphi=\omega t-kz$;
\begin{eqnarray}
\langle{J_j}\rangle =\frac{1}{2\pi}\int\limits_{-\varphi_\ast}^{\varphi_\ast}
 \left(-D_j\frac{\partial\wX_{s,j}}{\partial z}\right)d\varphi \qquad\qquad\nonumber\\[5pt]
{}+\frac{1}{2\pi}\int\limits_{\varphi_\ast}^{\varphi_\ast+2\pi}
 \left(-D_j\frac{\partial\wX_{\Sigma,j}}{\partial z}\right)d\varphi\,.
\label{eq-m10}
\end{eqnarray}

\section{Diffusion boundary layer}
\label{sec4}
For the case of single-component gas, the assumption of frozen profiles $X_{\Sigma,j}$ is accurate and fruitful~\cite{Goldobin-Krauzin-2015}, because the solubility profile is strictly dictated by the temperature and pressure fields and the diffusive transport downhill the solubility gradient is slow. No diffusion boundary layer appears near the surface, where the oscillating solute concentration is imposed. The case of two-component gas turns out to be essentially different, because the variation of the gas composition affects solubility and solute concentration profiles are not dictated solely by the temperature and pressure field. Indeed, in Fig.~(\ref{fig3}), one can clearly see the diffusion boundary layer with short-wave oscillations near the surface, which were never observed for a single-component gas. This boundary layer has to be taken into account and within this layer the profiles $X_{\Sigma,j}$ are not actually frozen. Although beyond the boundary layer these profiles can be assumed frozen, the diffusion boundary layer may affect the effective boundary conditions for the concentration fields within the zones of frozen profiles.

Let us consider the diffusion boundary layer. Since the diffusion boundary layer is localised near the surface on much shorter length scale than the scale of temperature wave, one may assume spatially uniform temperature field $\Theta(z,t)=\Theta_0\cos{\omega t}$ and neglect the hydrostatic pressure gradient.

The diffusive transport operates through solution and the diffusion coefficients are spatially uniform (for uniform temperature field);
\begin{equation}
\frac{\partial}{\partial t}\wX_{\Sigma,j}=D_j(\Theta)\frac{\partial^2}{\partial z^2}\wX_{s,j}\,.
\label{eq-bl01}
\end{equation}
Numerical simulations reveal that within the diffusion boundary layer with spatially uniform solubility field oscillating in time, the solute is undersaturated (bubbly phase disappears) only for a short part of the oscillation cycle, and this part vanishes as the ratio $D_j/\chi$ tends to zero. Hence, one can approximately assume the bubbly phase to be always-present and the solute concentration fields $\wX_{s,j}$ to obey Eqs.~(\ref{eq-m06}) and (\ref{eq-m07}).

Without hydrostatic pressure gradient Eq.~(\ref{eq-m06}) reads
\begin{equation}
K_{10}\widetilde{X}_{s,1}+K_{20}\widetilde{X}_{s,2}=-a_{12}\Theta\,,
\label{eq-bl02}
\end{equation}
where
\begin{equation}
a_{12}=Y_{10}a_1+Y_{20}a_2\,.
\label{eq-bl03}
\end{equation}
Eqs.~(\ref{eq-bl02}) and (\ref{eq-m07}) yield
\begin{eqnarray}
\mathcal{K}_0\wX_{s,1}=-Y_{10}a_{12}\Theta
 +K_{20}(Y_{20}\wX_{\Sigma,1}-Y_{10}\wX_{\Sigma,2})\,,
\label{eq-bl04}
\\[5pt]
\mathcal{K}_0\wX_{s,2}=-Y_{20}a_{12}\Theta
 -K_{10}(Y_{20}\wX_{\Sigma,1}-Y_{10}\wX_{\Sigma,2})\,,
\label{eq-bl05}
\end{eqnarray}
where
\begin{equation}
\mathcal{K}_0=Y_{10}K_{10}+Y_{20}K_{20}\,.
\label{eq-bl06}
\end{equation}
Substituting Eqs.~(\ref{eq-bl05}) and (\ref{eq-bl06}) into Eq.~(\ref{eq-bl01}) for $j=1,2$, one obtains
\begin{eqnarray}
\frac{\partial}{\partial t}\wX_{\Sigma,1}=\frac{D_{1}K_{20}}{\mathcal{K}_0}
\frac{\partial^2}{\partial z^2}(Y_{20}\wX_{\Sigma,1}-Y_{10}\wX_{\Sigma,2})\,,
\nonumber
\\[5pt]
\frac{\partial}{\partial t}\wX_{\Sigma,2}=-\frac{D_{2}K_{10}}{\mathcal{K}_0}
\frac{\partial^2}{\partial z^2}(Y_{20}\wX_{\Sigma,1}-Y_{10}\wX_{\Sigma,2})\,.
\nonumber
\end{eqnarray}
The latter equation system yields
\begin{equation}
\wX_{\Sigma,\mathrm{b.l.}}\equiv
 D_{2}K_{10}\wX_{\Sigma,1}+D_{1}K_{20}\wX_{\Sigma,2}=const\,,
\label{eq-bl07}
\end{equation}
\begin{equation}
\frac{\partial}{\partial t}\wX_{\Sigma,\mathrm{res}}
 =\mathcal{D}\frac{\partial^2}{\partial z^2}\wX_{\Sigma,\mathrm{res}}\,,
\label{eq-bl08}
\end{equation}
where
\begin{equation}
\wX_{\Sigma,\mathrm{res}}\equiv Y_{20}\wX_{\Sigma,1}-Y_{10}\wX_{\Sigma,2}\,,
\label{eq-bl09}
\end{equation}
\begin{equation}
\mathcal{D}\equiv
\frac{D_{2}K_{10}Y_{10}+D_{1}K_{20}Y_{20}}{\mathcal{K}_0}\,.
\label{eq-bl-D}
\end{equation}
The original variables can be calculated from $\wX_{\Sigma,\mathrm{b.l.}}$ and $\wX_{\Sigma,\mathrm{res}}$ as follows:
\[
\wX_{\Sigma,1}=\frac{Y_{10}\wX_{\Sigma,\mathrm{b.l.}}+D_1K_{20}\wX_{\Sigma,\mathrm{res}}}
  {D_{1}K_{20}Y_{20}+D_{2}K_{10}Y_{10}}\,,
\]
\[
\wX_{\Sigma,2}=\frac{Y_{20}\wX_{\Sigma,\mathrm{b.l.}}-D_2K_{10}\wX_{\Sigma,\mathrm{res}}}
  {D_{1}K_{20}Y_{20}+D_{2}K_{10}Y_{10}}\,.
\]

The solution to Eq.~(\ref{eq-bl08}) is a wave exponentially decaying with $z$ (which holds true as well for a time-dependent $\mathcal{D}(\Theta)$). Hence, beyond the diffusion boundary layer $\wX_{\Sigma,\mathrm{res}}=0$, and fields $\wX_{\Sigma,j}$ are determined by $\wX_{\Sigma,\mathrm{b.l.}}$. Calculating $\wX_{\Sigma,\mathrm{b.l.}}$ on the surface, where $\wX_{\Sigma,j}(0)=Y_{j0}a_j\Theta_0/K_{j0}$ (which corresponds to the maximal-over-period value of the solute concentration), one finds
\[
\wX_{\Sigma,\mathrm{b.l.}}=D_{2}Y_{10}a_{1}\Theta_0+D_{1}Y_{20}a_{2}\Theta_0\,,
\]
and {\it near the surface, immediately beyond the diffusion boundary layer}
\begin{equation}
\wX_{\Sigma,j}=\frac{Y_{j0}(D_{2}Y_{10}a_{1}\Theta_0+D_{1}Y_{20}a_{2}\Theta_0)}
 {D_{2}K_{10}Y_{10}+D_{1}K_{20}Y_{20}}\,.
\label{eq-bl10}
\end{equation}

Eq.~(\ref{eq-bl10}) provides effective boundary conditions at $z=0$ for the frozen-profile solutions outside the diffusion boundary layer.

\section{Beyond diffusion boundary layer}
\label{sec5}
\subsection{The case of small solubility oscillation amplitude ($a_j\Theta_0\ll1$) and bubbly horizon penetration depth ($kz\ll1$)}
For better understanding of the analytical solution it is convenient to consider the simplest case admitting purely analytical treatment. For this case we not only assume small oscillations of solubility and molecular diffusion coefficient but also take a note of the smallness of the penetration depth of the bubbly horizon for a small temperature oscillation amplitude, $e^{-kz}\approx1$.

To the leading order, Eq.~(\ref{eq-m10}) for $j=1$ yields
\begin{eqnarray}
\langle{J_1}\rangle =-D_{10}\frac{1}{2\pi}\int\limits_{-\varphi_\ast}^{\varphi_\ast}
  \frac{\partial\wX_{s,1}}{\partial z}d\varphi
\qquad\qquad\nonumber\\[5pt]
{} -D_{10}\left(1-\frac{\varphi_\ast}{\pi}\right)\frac{\partial\wX_{\Sigma,1}}{\partial z}\,,
\label{eq-s01}
\end{eqnarray}
where $D_{j0}=D_j(T_0)$. From Eqs.~(\ref{eq-m06})--(\ref{eq-m07}), one can find
\begin{eqnarray}
\mathcal{K}_0\wX_{s,1}=bY_{10}z-Y_{10}a_{12}\Theta
\qquad\qquad\nonumber\\[5pt]
{} +K_{20}(Y_{20}\wX_{\Sigma,1}-Y_{10}\wX_{\Sigma,2})\,.
\label{eq-s02}
\end{eqnarray}
For a steady solute distribution $\langle{J_j}\rangle=0$ and, thus, Eq.~(\ref{eq-s01}) with (\ref{eq-s02}) yields
\begin{eqnarray}
0=-D_{10}\left[\frac{\varphi_\ast}{\pi\mathcal{K}_0}\left(bY_{10}
 +K_{20}\left(Y_{20}\frac{d\wX_{\Sigma,1}}{dz}
\right.\right.\right.\qquad\nonumber\\[10pt]
\left.\left.
{}-Y_{10}\frac{d\wX_{\Sigma,2}}{dz}\right)\right)
+\frac{k a_{12}\Theta_0}{\pi}\frac{Y_{10}}{\mathcal{K}_0}\sin\varphi_\ast
\nonumber\\[10pt]
\left.
{}+\left(1-\frac{\varphi_\ast}{\pi}\right)\frac{d\wX_{\Sigma,1}}{dz}\right].
\label{eq-s05}
\end{eqnarray}
Simplifying the latter equation and performing similar derivation for specie 2, one can obtain
\begin{eqnarray}
Y_{10}\Bigg[\frac{\varphi_\ast}{\pi}\left(b
 -K_{10}\frac{d\wX_{\Sigma,1}}{dz}-K_{20}\frac{d\wX_{\Sigma,2}}{dz}\right)
\qquad
 \nonumber\\[10pt]
{}+\frac{k a_{12}\Theta_0}{\pi}\sin\varphi_\ast\Bigg]+\mathcal{K}_0\frac{d\wX_{\Sigma,1}}{dz}=0\,,
\label{eq-s06}
\\[10pt]
Y_{20}\Bigg[\frac{\varphi_\ast}{\pi}\left(b
 -K_{10}\frac{d\wX_{\Sigma,1}}{dz}-K_{20}\frac{d\wX_{\Sigma,2}}{dz}\right)
\qquad
 \nonumber\\[10pt]
{}+\frac{k a_{12}\Theta_0}{\pi}\sin\varphi_\ast\Bigg]+\mathcal{K}_0\frac{d\wX_{\Sigma,2}}{dz}=0\,.
\label{eq-s07}
\end{eqnarray}

\paragraph{Single-component gas:}
For the case of single-component gas, $Y_{20}=0$ and $\wX_{\Sigma,2}=0$, equation system (\ref{eq-s06})--(\ref{eq-s07}) reduces to
\begin{equation}
\varphi_\ast b + k a_1\Theta_0\sin\varphi_\ast+K_{10}(\pi-\varphi_\ast)\frac{d\wX_{\Sigma,1}}{dz}=0\,,
\label{eq-sg1}
\end{equation}
while Eq.~(\ref{eq-m08}) for $\varphi_\ast$ turns into
\begin{equation}
K_{10}\wX_{\Sigma,1}=bz-a_1\Theta_0\cos\varphi_\ast\,.
\label{eq-sg2}
\end{equation}
Employing relation (\ref{eq-sg2}), one can recast Eq.~(\ref{eq-sg1}) in terms of $\varphi_\ast$;
\[
\pi b+ k a_1\Theta_0\sin\varphi_\ast+(\pi-\varphi_\ast)a_1\Theta_0\sin\varphi_\ast\frac{d\varphi_\ast}{dz}=0\,.
\]
It is convenient to use dimensionless variables
\[
\xi=\frac{bz}{a_{12}\Theta_0}\,,\qquad
\varkappa=\frac{ka_{12}\Theta_0}{b}
\]
(in this case $a_{12}=a_1$). The dimensionless equation for $\varphi_\ast$ reads
\begin{equation}
\pi+\varkappa\sin\varphi_\ast+(\pi-\varphi_\ast)\sin\varphi_\ast\frac{d\varphi_\ast}{d\xi}=0\,.
\label{eq-sg3}
\end{equation}
The reference values of dimensionless $\xi$ is of the order of magnitude of $1$; therefore, the assumption $kz=\varkappa\xi\ll1$ requires $\xi\ll1$. Hence, the second term in Eq.~(\ref{eq-sg2}) should be neglected; one finds
\begin{equation}
\pi+(\pi-\varphi_\ast)\sin\varphi_\ast\frac{d\varphi_\ast}{d\xi}=0\,,
\label{eq-sg4}
\end{equation}
which is identical to Eq.~(19) in \cite{Goldobin-Krauzin-2015}, and can be integrated with the initial condition $\varphi_\ast(\xi=0)=\pi$;
\begin{equation}
(\pi-\varphi_\ast)\cos\varphi_\ast+\sin\varphi_\ast=\pi\xi\,.
\label{eq-sg5}
\end{equation}
Eq.~(\ref{eq-sg5}) provides an implicit dependence of $\varphi_\ast(\xi)$; $\varphi_\ast(\xi)$ decreases monotonously with depth $\xi$ from $\varphi_\ast(0)=\pi$ till $\varphi_\ast(1)=0$. The bubbly horizon penetration depth is $\xi_b=1$. With known $\varphi_\ast\big(bz/(a_1\Theta_0)\big)$, one can employ Eq.~(\ref{eq-sg2}) to calculate $\wX_{\Sigma,1}(z)$.

Beneath the penetration depth of the bubbly horizon $\xi_b=1$, the solute concentration is spatially uniform and constant in time; to the leading order,
\begin{equation}
\wX_{\Sigma,1}(\infty)|_{Y_{20}=0}=0\,.
\label{eq-sg6}
\end{equation}

\paragraph{Two-component gas:}
As clearly shown for a single-component case, the consistency of approximation $kz\ll1$ suggests to neglect the term $ka_{12}\Theta_0\sin\varphi_\ast$ in Eqs.~(\ref{eq-s06}) and (\ref{eq-s07}). The sum of Eq.~(\ref{eq-m06}) multiplied by $K_{10}$ and Eq.~(\ref{eq-m07}) multiplied by $K_{20}$ yields
\begin{equation}
\frac{\varphi_\ast}{\pi}\left(b-\frac{dZ}{dz}\right) +\frac{dZ}{dz}=0\,,
\label{eq-tg1}
\end{equation}
where
\begin{equation}
Z=K_{10}\wX_{\Sigma,1}+K_{20}\wX_{\Sigma,2}\,.
\label{eq-tg2}
\end{equation}
Eq.~(\ref{eq-m08}) in terms of $Z$ reads
\begin{equation}
Z=bz-a_{12}\Theta_0\cos\varphi_\ast\,.
\label{eq-tg3}
\end{equation}

Substitution of $Z$ from Eq.~(\ref{eq-tg3}) into Eq.~(\ref{eq-tg1}) yields in terms of $\varphi_\ast(\xi)$ an equation identical to Eq.~(\ref{eq-sg4}). However, it should be integrated with boundary conditions accounting for the diffusion boundary layer, Eq.~(\ref{eq-bl10});
\[
Z(0)=\mathcal{K}_0\frac{D_{2}Y_{10}a_{1}\Theta_0+D_{1}Y_{20}a_{2}\Theta_0}
 {D_{2}K_{10}Y_{10}+D_{1}K_{20}Y_{20}}\,,
\]
and $\varphi_\ast(0)$ does not equal $\pi$, as for the single-component case, but [{\it cf.} Eq.~(\ref{eq-tg3})]
\begin{eqnarray}
\cos\varphi_\ast(0)=-\frac{Y_{10}K_{10}+Y_{20}K_{20}}{Y_{10}a_1+Y_{20}a_2}
\qquad\qquad
\nonumber\\[7pt]
\times\frac{D_{2}Y_{10}a_{1}+D_{1}Y_{20}a_{2}}
 {D_{2}K_{10}Y_{10}+D_{1}K_{20}Y_{20}}\,.
\label{eq-tg4}
\end{eqnarray}
Instead of relation (\ref{eq-sg5}), one finds
\begin{equation}
(\pi-\varphi_\ast)\cos\varphi_\ast+\sin\varphi_\ast=\pi(\xi+1-\xi_b)\,,
\label{eq-tg5}
\end{equation}
where the penetration depth of the bubbly horizon
\[
\xi_b=1-\frac{[\pi-\varphi_\ast(0)]\cos\varphi_\ast(0)+\sin\varphi_\ast(0)}{\pi}
\]
is decreased as compared to the case of a single-component gas.

The difference of Eq.~(\ref{eq-m06}) multiplied by $Y_{20}$ and Eq.~(\ref{eq-m07}) multiplied by $Y_{10}$ yields
\[
\frac{d}{dz}\wX_{\Sigma,\mathrm{res}}=0\,,
\]
where $\wX_{\Sigma,\mathrm{res}}$ is determined by Eq.~(\ref{eq-bl09}). With the boundary conditions (\ref{eq-bl10}) one obtains $\wX_{\Sigma,\mathrm{res}}=0$.

The specie distributions can be calculated from $Z$ and $\wX_{\Sigma,\mathrm{res}}$;
\begin{eqnarray}
\wX_{\Sigma,1}=\frac{Y_{10}Z+K_{20}\wX_{\Sigma,\mathrm{res}}}{\mathcal{K}_0}
\qquad\qquad\qquad
\label{eq-tg6}\\
=\frac{Y_{10}}{\mathcal{K}_0}\Bigg(bz-a_{12}\Theta_0\cos\varphi_\ast\Bigg)\,,
\label{eq-tg7}\\[5pt]
\wX_{\Sigma,2}=\frac{Y_{20}Z-K_{10}\wX_{\Sigma,\mathrm{res}}}{\mathcal{K}_0}
\qquad\qquad\qquad
\label{eq-tg8}\\
=\frac{Y_{20}}{\mathcal{K}_0}\Bigg(bz-a_{12}\Theta_0\cos\varphi_\ast\Bigg)\,.
\label{eq-tg9}
\end{eqnarray}

Beneath the penetration depth of the bubbly horizon $\xi_b$, where $\varphi_\ast=0$, the solute concentration is spatially uniform and constant in time; to the leading order,
\[
\wX_{\Sigma,1\infty}=\wX_{\Sigma,2\infty}=0\,,
\]
meaning that the composition of solution is not changed compared to the case of no temperature oscillation.

To summarize the consideration of this subsection, in terms of the essential quantifier $\varphi_\ast$ the case of a two-component atmosphere is similar to the case of a single-component atmosphere with effective parameter $a_{12}$ instead of $a_j$ and $\mathcal{K}_0$ instead of $K_{j0}$. However, the diffusion boundary layer effectively reduces the most upper part of the bubbly horizon. Indeed, for the single-component case $\varphi_\ast(0)=\pi$, while for the two-component case $\varphi_\ast(0)$ is determined by Eq.~(\ref{eq-tg4}), {\it i.e.}, the profile $\varphi_\ast(\xi)$ for the latter case is the profile for the former case shifted towards the surface.

Noteworthy, the constructed analytical theory is an approximation but not a rigorous limiting case. The analytical theory requires small $\Theta_0$. Meanwhile, for small $\Theta_0$ the penetration depth of the bubbly zone is small~\footnote{From the definition of $\xi$, one can see that the penetration depth is a linear function of $\Theta_0$} and can become commensurable with the thickness of the diffusion boundary layer
\[
\delta_{\mathrm{diff}}=\sqrt{2\mathcal{D}/\omega}\;.
\]
In the latter case the approximation of `frozen' profiles $X_{\Sigma,j}(z)$ is invalid. Thus, the frozen profile approximation is not compatible with the limit of vanishing $\Theta_0$. Nonetheless, for moderately small $\Theta_0$, both approximations can be satisfactory accurate.

\subsection{The case of moderate penetration depth of the bubbly horizon}
The analytical theory constructed for the case of a small penetration depth provides opportunity of a purely analytical solution, significantly benefits the understanding of the system dynamics, and provides assessment on characteristic features of the system, such as relation between the penetration depth and temperature oscillation amplitude $\Theta_0$. With this basic theoretical picture of the system, one can tackle the task of constructing the theory for the case of moderate penetration depth, where $kz$ (or $\varkappa\xi$) is non-small within the bubbly zone.

For this case, calculation of the average diffusion fluxes (\ref{eq-m10}) requires account for the dependence of the diffusion coefficients on temperature;
\begin{equation}
D_j(T)=D_{j0}\big(1+\gamma_j\Theta+\mathcal{O}\big((\gamma_j\Theta)^2\big)\big)\,.
\label{eq-c01}
\end{equation}
After laborious but straightforward calculations one can obtain an amended version of equation system (\ref{eq-s06})--(\ref{eq-s07});
\begin{widetext}
\begin{eqnarray}
&&Y_{10}\Bigg[\frac{(\varphi_\ast+\gamma_1\Theta_0e^{-kz}\sin\varphi_\ast)}{\pi}\left(b
 -{\textstyle\sum\limits_j}K_{j0}\frac{d\wX_{\Sigma,j}}{dz}\right)
 \nonumber\\[5pt]
&&\qquad\qquad\qquad
 {}+\frac{k a_{12}}{\pi}\bigg(\Theta_0e^{-kz}\sin\varphi_\ast
 +\frac{\gamma_1\Theta_0^2}{2}e^{-2kz}\Big(\varphi_\ast-\frac12\sin{2\varphi_\ast}\Big)\bigg)\Bigg]
 +\mathcal{K}_0\frac{d\wX_{\Sigma,1}}{dz}=0\,,
\label{eq-c02}
\\[10pt]
&&Y_{20}\Bigg[\frac{(\varphi_\ast+\gamma_2\Theta_0e^{-kz}\sin\varphi_\ast)}{\pi}\left(b
 -{\textstyle\sum\limits_j}K_{j0}\frac{d\wX_{\Sigma,j}}{dz}\right)\qquad
 \nonumber\\[5pt]
&&\qquad\qquad\qquad
 {}+\frac{k a_{12}}{\pi}\bigg(\Theta_0e^{-kz}\sin\varphi_\ast
 +\frac{\gamma_2\Theta_0^2}{2}e^{-2kz}\Big(\varphi_\ast-\frac12\sin{2\varphi_\ast}\Big)\bigg)\Bigg]
+\mathcal{K}_0\frac{d\wX_{\Sigma,2}}{dz}=0\,.
\label{eq-c03}
\end{eqnarray}

The sum of Eq.~(\ref{eq-c02}) multiplied by $K_{10}$ and Eq.~(\ref{eq-c03}) multiplied by $K_{20}$ yields
\begin{equation}
\frac{(\varphi_\ast+\gamma_{12}\Theta_0e^{-kz}\sin\varphi_\ast)}{\pi}\left(b-\frac{dZ}{dz}\right)
 +\frac{k a_{12}}{\pi}\bigg(\Theta_0 e^{-kz}\sin\varphi_\ast
 +\frac{\gamma_{12}\Theta_0^2}{2}e^{-2kz}\Big(\varphi_\ast-\frac12\sin{2\varphi_\ast}\Big)\bigg)
 +\frac{dZ}{dz}=0\,,
\label{eq-c04}
\end{equation}
where
\begin{equation}
\gamma_{12}=\frac{\gamma_1Y_{10}K_{10}+\gamma_2Y_{20}K_{20}}{Y_{10}K_{10}+Y_{20}K_{20}}\,.
\label{eq-c05}
\end{equation}
Eq.~(\ref{eq-m08}) for $\varphi_\ast$ yields
\begin{equation}
Z=bz-a_{12}\Theta_0e^{-kz}\cos\varphi_\ast\,.
\label{eq-c06}
\end{equation}

Substituting Eq.~(\ref{eq-c06}), one can recast Eq.~(\ref{eq-c04}) in a dimensionless form for $F=e^{-\varkappa\xi}\cos\varphi_\ast$ (notice also the relation $Z=bz-a_{12}\Theta_0F$);
\begin{equation}
\frac{d\xi}{dF}=\frac{\pi-\arccos(Fe^{\varkappa\xi})-\gamma_{12}\Theta_0\sqrt{e^{-2\varkappa\xi}-F^2}}
 {\pi+\frac12\varkappa\gamma_{12}\Theta_0e^{-2\varkappa\xi}\arccos(Fe^{\varkappa\xi})
 +\varkappa\big(1-\frac12\gamma_{12}\Theta_0F\big)\sqrt{e^{-2\varkappa\xi}-F^2}}\,.
\label{eq-c07}
\end{equation}
\end{widetext}
Eq.~(\ref{eq-c07}) should be integrated from the initial condition $\xi\big(F=\cos\varphi_\ast(0)\big)=0$ (at the surface) till the point where the condition $F=e^{-\varkappa\xi}$ will be fulfilled (at the base of the bubbly horizon); $\cos\varphi_\ast(0)$ is determined by Eq.~(\ref{eq-tg4}). We integrate $(d\xi/dF)$ instead of $(dF/d\xi)$ on purpose, as it allows an easy handling of singularities $dF/d\xi=\infty$ at the surface and at the base of the bubbly horizon.

\begin{figure}[!t]
\centerline{
\includegraphics[width=0.375\textwidth]%
 {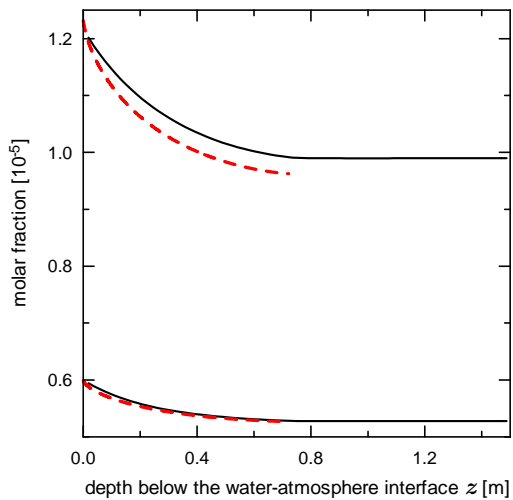}
}

  \caption{(Color online) The theoretical profiles (red dashed lines) are compared to the average profiles of $X_{\Sigma,j}$ from numerical simulation (black solid lines) for nitrogen (top) and oxygen (bottom). Remarkably the difference between the average profiles from numerical simulations for $D_j=D_{\mathrm{mol},j}$ and $D_j=0.01D_{\mathrm{mol},j}$ is less than the line thickness (only the former profile is plotted).
  Parameters: $T=300\,\mathrm{K}$, $\Theta_0=15\,\mathrm{K}$.
 }
  \label{fig5}
\end{figure}

Let us now calculate the quantifiers of composition of the solution. Similarly to the case of small penetration depth, the difference of Eq.~(\ref{eq-c02}) multiplied by $Y_{20}$ and Eq.~(\ref{eq-c03}) multiplied by $Y_{10}$ yields the differential equation for $\wX_{\Sigma,\mathrm{res}}=Y_{20}\wX_{\Sigma,1}-Y_{10}\wX_{\Sigma,2}$;
\begin{eqnarray}
\frac{d}{dz}\wX_{\Sigma,\mathrm{res}}=\frac{Y_{10}Y_{20}}{\pi}(\gamma_1-\gamma_2)a_{12}\Theta_0^2
\qquad\qquad
\nonumber\\[8pt]
 \times\bigg[-e^{-kz}\sin\varphi_\ast\frac{d}{dz}(e^{-kz}\cos\varphi_\ast)
\qquad
\nonumber\\[5pt]
 {}-\frac{k}{2}e^{-2kz}\Big(\varphi_\ast-\frac12\sin{2\varphi_\ast}\Big)\bigg].
\nonumber
\end{eqnarray}
The latter equation can be recast in a dimensionless form convenient for integration along with Eq.~(\ref{eq-c07});
\begin{eqnarray}
&&d\wX_{\Sigma,\mathrm{res}}=\frac{Y_{10}Y_{20}}{\pi}(\gamma_1-\gamma_2)a_{12}\Theta_0^2
\nonumber\\[8pt]
&&\qquad\qquad
 \times\bigg[-\sqrt{e^{-2\varkappa\xi}-F^2}\,dF
\label{eq-c08}\\[5pt]
&& {}-\frac{\varkappa}{2}\left(e^{-2\varkappa\xi}\arccos(e^{\varkappa\xi}F)
   -F\sqrt{e^{-2\varkappa\xi}-F^2}\right)d\xi\bigg].
\nonumber
\end{eqnarray}
The value of $\wX_{\Sigma,\mathrm{res}}$ at the surface is determined by boundary conditions (\ref{eq-bl10});
\begin{equation}
\left.\wX_{\Sigma,\mathrm{res}}\right|_{z=0}=0\,,
\label{eq-c09}
\end{equation}
which serves as the initial condition for integration of Eq.~(\ref{eq-c08}). With known $Z$ and $\wX_{\Sigma,\mathrm{res}}$, one can evaluate
\begin{eqnarray}
\wX_{\Sigma,1}=\frac{Y_{10}Z+K_{20}\wX_{\Sigma,\mathrm{res}}}{\mathcal{K}_0}\,,
\label{eq-c10}\\
\wX_{\Sigma,2}=\frac{Y_{20}Z-K_{10}\wX_{\Sigma,\mathrm{res}}}{\mathcal{K}_0}\,.
\label{eq-c11}
\end{eqnarray}
In Fig.~\ref{fig5}, one can appreciate the agreement between the theory (\ref{eq-c07})--(\ref{eq-c11}) with initial condition (\ref{eq-tg4}) and the results of numerical simulations.

%%%%%%%%%%%%%%%%%%%%%%%%%%%%%%%%%%%%%%%%%%%%%%%%%%%%%%%%%
%%%%%%%%%%%%%%%%%%%%%%%%%%%%%%%%%%%%%%%%%%%%%%%%%%%%%%%%%

Note, $\wX_{\Sigma,\mathrm{res}}$ [Eqs.~(\ref{eq-c08})--(\ref{eq-c09})] is of higher order of smallness in $\Theta_0$ than $Z$. Hence, it is small for small oscillations of solubility. Moreover, it is also proportional to the difference in the the temperature dependence of the diffusion coefficients, $(\gamma_1-\gamma_2)$; without difference $(\gamma_1-\gamma_2)$, discrepancies in $D_{j0}$ or in solubility properties of species cannot create variation of $\wX_{\Sigma,\mathrm{res}}$ across the bubbly horizon. To the leading order in $\Theta_0$, the solution composition beneath the bubbly horizon is
\begin{equation}
\wX_{\Sigma,1\infty}\approx\wX_{\Sigma,2\infty}
\approx0\,,
\label{eq-c12}
\end{equation}
meaning that the composition of solution is not changed as compared to the case of no temperature oscillation.

\section{Conclusion}
\label{sec6}
We have studied the effect of surface temperature oscillations on the infiltration of a weakly soluble substance into a liquid-saturated porous medium. Bearing in mind the problem of the saturation of sediments with the atmospheric gases under the conditions of annual or daily surface temperature oscillations and other possible geological systems subject to cyclic thermal conditions, where the guest substances are rarely single-component, we considered the case of a two-components substance ({\it e.g.}, nitrogen+oxygen for the atmosphere).
Specifically, we assumed the liquid-saturated porous half-space contacting with a reservoir of a weakly soluble substance. Temperature of the contact interface was assumed to oscillate sinusoidally. The interface temperature oscillation creates the temperature wave propagating into the porous medium and decaying with depth.
The solubility wave, associated with the temperature wave, creates time-dependent spatial intermittency between the zones of nondissolved phase and the zones of undersaturated solution.

Because of the smallness of ratio $D/\chi$, which is $\sim10^{-3}$ for typical liquids, the diffusion transfer in the system is much slower than the temperature (and related solubility) variation. As a result, the profile of the net molar fraction of the guest molecules in pores, $X_\Sigma$ (``net'' means ``solute+nondissolved phase''), is almost frozen during one oscillation cycle. For gases, the profile was shown to attain the maximal-over-period solubility near the surface, monotonously decays with depth within the zone where the nondissolved phase can be observed---so called, ``bubbly horizon''---and becomes nearly constant in space and time beneath the horizon.

From the view point of physics, the appearance of a diffusion boundary layer reported for multicomponent substances is of interest. For single-component guest substances this boundary layer never appears since in this thin near-surface zone the concentration profile is the solubility one, which is unambiguously dictated by the temperature field~\cite{Goldobin-Krauzin-2015}. For multicomponent substances, the solubility depends on the fraction of components in the nondissolved phase. Imposed nonstationarity of the concentration (solubility) at the surface forces the diffusive redistribution of guest molecules of two sorts in the porous medium. The wave of this redistribution processes creates the wave of solubility on the same spatial-temporal scales determined by the effective diffusion coefficient $\mathcal{D}$ (\ref{eq-bl-D}).
With wave decays within a transient zone --- the diffusion boundary layer of thickness $\delta_\mathrm{diff}=\sqrt{2\mathcal{D}/\omega}$. We have show that beyond this boundary layer an effective boundary condition can be adopted for the composition of the guest substance (\ref{eq-bl10}), and the gas transport is equivalent to a single-component one with effective parameters of solubility $\mathcal{K}_0$ and its temperature dependence $a_{12}$, (\ref{eq-bl06}) and (\ref{eq-bl03}).

The boundary layer is thin but it effectively reduces the capacity of the bubbly horizon: within the bubbly horizon the part of the cycle when the nondissolved phase is present, $\varphi_\ast/\pi$, monotonously decreases from $1$ at the surface to $0$ at the base horizon. Without the diffusion boundary layer, $\varphi_\ast(z=0)=\pi$, while in its presence, immediately beyond the layer, $\varphi_\ast(z=+0)$ is lowered from $\pi$ by a finite value. The lowering is stronger for a stronger dissimilarity in the diffusive mobilities and the temperature dependencies of species solubilities; indeed, Eq.~(\ref{eq-tg4}) yields the minimal value $\cos\varphi_\ast(0)=-1$ for $D_1=D_2$ or $K_{10}/K_{20}=a_1/a_2$. Approximately, this corresponds to the shift of profiles plotted in Fig.~\ref{fig5} leftwards by a value $\propto[1+\cos\varphi_\ast(0)]$, leaving smaller integral profile excesses above their asymptotic values $X_{\Sigma,1\infty}$ and $X_{\Sigma,2\infty}$.

Noteworthy, within the diffusion boundary layer the hydrostatic pressure variation is negligible; therefore, the theory we have constructed for it is equally applicable to the cases of solid and liquid nondissolved phases. However, the theory for the bulk of the bubbly horizon is heavily affected by the hydrostatic pressure gradient. The generalization of this theory to solids/liquids requires the development of an alternative version of the theory in Sec.~\ref{sec5}, which is beyond the scope of this paper.
For high-frequency temperature oscillations the temperature wave penetration depth is small and the gas solubility profile is not affected by the hydrostatic pressure trend up to this depth. This case will be also mathematically equivalent to the case of solid/liquid nondissolved phase.

\acknowledgements{
The work has been supported by the Ministry of Science and Higher Education of the Russian Federation (theme no.\ 121112200078-7).
}

\end{document}